\documentclass[showpacs,aps,twocolumn]{revtex4}
\usepackage{graphicx}
\usepackage{amsmath,amssymb,amsfonts}
\usepackage{array}
\usepackage{url}
\usepackage{hyperref}
\usepackage{multirow}
\usepackage{float}
\usepackage{lineno}
\usepackage{xspace}
\usepackage{ulem}
\begin{document}

\title{Thermal Evolution and Axion Emission Properties of Strongly Magnetized Neutron Stars}
\author{Shubham Yadav}
\email{shubhamphy28@gmail.com}

\author{M. Mishra}
\email{madhukar@pilani.bits-pilani.ac.in}

\author{Tapomoy Guha Sarkar}
\email{tapomoy@pilani.bits-pilani.ac.in}

\affiliation{Department of Physics, Birla Institute of Technology and Science, Pilani, Rajasthan, India}

\author{Captain R. Singh}
\email{captainriturajsingh@gmail.com}
\affiliation{Department of Physics, Indian Institute of Technology Indore, Simrol, Indore 453552, India}

\begin{abstract}
Emission properties of compact astrophysical objects such as Neutron stars (NSs) are associated with crucial astronomical observables. In the current work, we obtain the mass, pressure profiles of the non-rotating NSs using the modified Tolman Oppenheimer Volkoff (TOV) system of equations in the presence of intense magnetic field. We obtain the profiles by using a specific distance-dependent magnetic field in the modified TOV equations. We employ three different equations of states (EoS) to solve the TOV equations by assuming the core of NSs comprises a hadronic matter. Employing the above profiles, we determine the cooling rates of spherically symmetric NSs as a function of time with and without including the magnetic field using the NSCool code. We have also determined the cooling rates as a function of radius for three different NSs. 
Furthermore, we determine the luminosity of neutrinos, axions, and photons emitting from the NSs in the presence and absence of a magnetic field for an axion mass $16$ meV and three different EoS. Our comparative study indicates that the cooling rate and luminosities of neutrinos, axions, and photons change significantly due to the impact of the strong magnetic field. We also find that due to the magnetic field, the axion mass bound increases slightly compared to without a magnetic field.

\end{abstract}
\date{\today}
\maketitle

\section{Introduction}
\label{intro}
Neutron Stars (NSs)~\cite{doi:10.1146/annurev-astro-081915-023329} are considered crucial observables as they allow testing physical theories at very high central matter density.
Recent observations of gravitational wave events involving mergers of NSs have opened up new frontiers in the area of NSs research~\cite {2006ARNPS56327P,Page_2004,PAGE2006497,1998nspt.conf..183P,10.1093/ptep/ptab099,2019A&A...629A..88P,Potekhin:2014hja,2020ApJ...888...97B,2003A&A...407..265Y,PhysRevC.98.035802}. The detailed study of NSs physics is also crucial to make a comparative study with futuristic collider experiments like the Compressed Baryonic Matter Experiment (CBM) involving matter at very high density and low temperatures.\\
\noindent Recent investigations have revealed new information about the potency of magnetic fields inside and on the surface of compact objects~\cite{Dexheimer:2017fhy, DEXHEIMER2017487, 2019PhRvC..99e5811C,2015MNRAS.447.1598B}. The macroscopic structure, evolution, and observable astrophysical properties of NSs are strongly determined by the equation of state (EoS) of the matter and other sources in the energy-momentum tensor. Several observations such as, Soft Gamma Ray repeaters and X-ray pulsars indicate that a specific group of NSs, called magnetars~\cite{dexheimer2012hybrid},  possess a very strong magnetic fields. In the literature, it is established that the internal structure of NSs and their cooling properties get affected by the distribution of strong magnetic field in the interior~\cite {PhysRevD.37.2042,doi:10.1146/annurev.astro.42.053102.134013}. Magnetars are strongly magnetized NSs that exhibit a wide array of X-ray activity, with anticipated magnetic field intensities of up to $10^{18}$ Gauss~\cite{doi:10.1146/annurev-astro-081915-023329}. This magnetic field is much larger than the level at which the electron's Landau level energies match its rest mass. In fact, the Virial theorem predicts the inner magnetic fields to be as high as $10^{18}$ Gauss. The exact origin of magnetic fields in different astrophysical objects is still an open area of research~\cite{Beznogov_2023,2012A&A...538A.115P}.
Magnetars have attracted the attention of scientists due to their extraordinarily strong magnetic fields dependent properties~\cite{Gomes_2017}. This magnetic field may significantly modify the emission of high-energy electromagnetic radiation from these objects and thereby affect their cooling properties~\cite{Beznogov_2023,Yakovlev_2005,PhysRevD.37.2042,PAGE2006497,YAKOVLEV2004523,2015SSRv..191..239P,refId04,Jonker_2007}.There is also a renewed interest in studying NSs as they have been proposed to be the source of axionic DM particles.\\

Even after several years of searches,the exact bound for axion mass (and coupling constant) is still not well established. The axion mass bound from various experiments are as follows; MADMAX and plasma haloscopes~\cite{brun2019new,PhysRevLett.123.141802} provide axion mass in the range $\sim$ $40–400$ $\mu$ eV, ADMX and HAYSTAC give the axion mass $m_{a}\sim 1–100$ $\mu$eV, DM-Radio~\cite{7750582} and CASPEr~\cite{PhysRevX.4.021030,PhysRevLett.126.141802,Garcon_2018} and ABRACADABRA~\cite{PhysRevLett.117.141801,PhysRevLett.122.121802,PhysRevLett.127.081801}, indicate the axion masses $m_{a} \ll$  $\mu$eV. 
The ADMX experiment has gained sensitivity to Dean-Fischler-Srednitsky-Zhitnitsky (DFSZ) QCD axion model with axion mass range $m_{a} \sim$ 2.7 $-$ 4.2 $\mu$eV.~\cite{PhysRevLett.120.151301,PhysRevLett.124.101303,PhysRevLett.127.261803,DINE1981199,Zhitnitsky:1980tq}.However, Blackhole superradiance~\cite{PhysRevD.81.123530,PhysRevD.83.044026} indicates QCD axion masses $> 2\times 10^{-11}$ eV.
In the context of Kim-Shifman-Weinstein-Zakharov (KSVZ) axion model, Buschmann et al.\cite{Buschmann:2021juv} restrict axion mass at $m_{a}\leqslant 16$ meV ($95 \%$ confidence level) and finds no evidence for axions in the isolated NSs data.
The ADMX~\cite{PhysRevD.64.092003} and HAYSTAC~\cite{backes2021quantum} experimental results in context to strongly coupled KSVZ model~\cite{PhysRevLett.43.103,SHIFMAN1980493} reveals that there is roughly for $\sim$ ten orders of magnitude of parameter space for the axion mass still remains unprobed.
Axionic DM, with mass not in meV range is challenging to explore in the laboratory.~\cite{Armengaud_2019,Liu_2022,Co_2020,10.21468/SciPostPhys.10.2.050,arvanitaki2014resonantly,carosi2020microwave}. Recent work~\cite{PhysRevD.107.103017,PhysRevD.109.023001}, gives the bound on axion mass as $m_{a}\leqslant$ 10 meV.
\noindent In the current work, we have investigated the impact of magnetic field on the pressure and mass profile of the non-rotating NSs with mass $M \sim 1.4M_{\odot}$. We employ three microscopic model-based EoS: APR, FPS, and SLY. The distance-dependent magnetic field is calculated using an eighth-order polynomial with a high value of the central magnetic field. Thus, the magnetic field obtained is used in a modified TOV system of equations to find the pressure and mass profiles. We have studied the impact of a strong magnetic field on the cooling properties of NSs for different EoS using the NSCool code~\cite {2016ascl.soft09009P}. We have also investigated the role of a strong magnetic field on the luminosities of observables (neutrinos, axions, and photons). We conclude that the magnetic field affects the luminosities through a change in the thermal profile of the star.\\

\noindent The paper is organized as follows. In Section~\ref{intro}, we present a brief Introduction. Section~\ref{form} describes the modified TOV equations, Equation of states, NSs cooling, and QCD Axions. We have also subsequently discussed the axion and neutrino emission rates. In Section~\ref{result}, we present the discussions on the results of the impact of magnetic fields on various observables and macroscopic NS properties. We have discussed how including a magnetic field affects the profiles and cooling rates. Further, we also investigate the luminosities (neutrinos, axions, and photons) for an axion mass $16$ meV. Finally, in Section \ref{conc}, we summarize and conclude the work. 
\section{Formalism}
\label{form}

\subsection{TOV Equations for non-rotating Neutron star}
The present investigation is based on the modified TOV system of equations in the presence of a strong magnetic field. A profile for the magnetic field's strength is required to analyze the impact of magnetic fields for a particular EoSs of NSs.
In Schwarzschild coordinates, the space-time metric inside the spherically symmetric star is given by:

\begin{equation}
ds^2=-\exp(2\phi )c^2dt^2+\frac{dr^2}{1-\frac{2G\,m(r)}{rc^2}}+r^2d\Omega,
\end{equation}
where, $\exp(2\phi (R))=1-\frac{2GM}{Rc^2}$.\\

The interior solution of Einstein's equation of General theory of relativity for this metric yields the following first-order differential equations (TOV)~\cite{PhysRev.55.364,PhysRev.55.374}: 
\begin{equation}
\frac{dm}{dr}=4\pi r^2 \epsilon(r)
\end{equation}
\begin{equation}
\frac{dP}{dr}=-\frac{G\left(\epsilon   +\frac{P}{c^2}\right)\left(m(r)\,c^2 + 4 \pi r^3 P\right)}{r\left(r\,c^2-2Gm(r)\right)}
\end{equation}

The above equations are to be integrated to obtain the pressure $P(r)$ and mass $m(r)$ inside the interior at a distance $r$ under hydrostatic conditions.
The boundary conditions for mass are $m(r = 0) = 0$, $m(r = R) = M$. The core density is $\epsilon(r=0) = \epsilon_0$ and core pressure $P(r = 0) = P_0$. The pressure decreases towards the surface, and at the radius, $R$ of the star, the pressure $P(r= R)$ drops to zero~\cite{1968ApJ...153..807H}.
Under the effect of the magnetic field, a magnetic energy density term is included in the matter density term of the TOV system of equations. Strong magnetic fields in magnetars will contribute as a source of the energy-momentum tensor and consequently affect the interior solutions. It has been shown that the TOV equations~\cite{PhysRev.55.374} may be modified by accounting for the contribution from the magnetic field to the energy density and a Lorentz force term under the condition of hydrostatic equation ~\cite {2019PhRvC..99e5811C}.\\
The modified TOV equations in the presence of a magnetic field are given by: 
\begin{equation}
\frac{dm}{dr}=4\pi r^{2}\left ( \epsilon +\frac{B^{2}}{2\mu _{0}c^2} \right )
\end{equation}

\begin{equation}
\frac{d\phi }{dr}=\frac{G\left ( m(r)+  4 \pi r ^3  P/c^{2} \right )}{r\left ( r\,c^2-2G\,m(r)\right )}
\end{equation}

\begin{equation}
\frac{dP}{dr}=-c^{2} \left ( \epsilon +\frac{B^2}{2\mu _{0}c^2}+\frac{P}{c^{2}} \right )\left ( \frac{d\phi }{dr}-{\mathcal L}\left ( r \right ) \right ),
\end{equation}
where $\mathcal L(r)$  denotes a  Lorentz force contribution~\cite {2019PhRvC..99e5811C}.
The phenomenological modeling of the effect of the magnetic field~\cite{10.1093/mnras/sty776} is used to write an effective Lorentz force term in the modified TOV equations as~ \cite {2019PhRvC..99e5811C};
\begin{equation}
\frac{{\mathcal L}\left ( r \right )}{ 10^{-41}} = B_{c}^2\left[-3.8 \left(\frac{r}{\bar r} \right ) + 8.1 \left(\frac{r}{\bar r} \right )^3  -1.6   \left(\frac{r}{\bar r} \right )^5 -2.3   \left(\frac{r}{\bar r} \right )^7 \right ]
\end{equation}

The distribution of the magnetic field in the interior of the star shall affect the mass and pressure profiles.
Several such magnetic field profiles are available in the literature to study the structure of a NSs in strong magnetic fields ~\cite{Dexheimer:2017fhy,DEXHEIMER2017487,2019PhRvC..99e5811C,Lopes_2015,2013arXiv1307.7450G}.

Chatterjee et al.~\cite{2019PhRvC..99e5811C} reported a thorough analysis of the magnetic field configurations in non-rotating NSs using an axisymmetric numerical code, altering the mass, the strength of the magnetic field, and the equation of state. \\
  
\noindent Apart from the magnetic field distribution in a NSs~\cite{dexheimer2012hybrid,Reddy:2021rln,1995A&A...301..757B}, the EoS of matter under extreme conditions of very high density determines the internal structure and cooling properties of NSs~\cite {PhysRevD.37.2042,PAGE2006497,refId0,YAKOVLEV2004523,10.1093/mnras/sty776,10.1093/mnras/stx366}.  
  
\noindent Several groups have attempted to compute the EoS of the matter contained inside the core of NSs with and without considering finite baryon chemical potential~\cite{Dexheimer:2017fhy,DEXHEIMER2017487,2019PhRvC..99e5811C,Page_2004}. The study of magnetic field distribution as a function of distance, baryon chemical potential, rotation frequency, etc., has been presented in the Refs.~\cite{DEXHEIMER2017487,2019PhRvC..99e5811C}.\\
The EoS of Strongly magnetized NSs substantially affect their internal structure and macroscopic properties like mass and radius~\cite{PRAKASH19971,Lattimer_2001,Page_2004}. 
However, for stable equilibrium, a NSs must contain both poloidal (i.e., having only components $B_r$ and $B_\theta$)  and toroidal (i.e., having only the azimuthal component $B_\phi$) components since both are unstable on their own \cite{1956ApJ...123..498P,2006A&A...450.1077B,refId0}. 
It is possible that such stars become unstable unless the toroidal and poloidal fields are both present and in some ratio~\cite{10.1111/j.1365-2966.2008.14034.x}.
Further, relativistic considerations of such highly magnetized NSs show that the magnetic energy decays on a timescale which is a function of the Alfvén crossing time and the rotation speed. It is relatively short compared to any evolutionary timescale~\cite{2004Natur.431..819B,1995A&A...301..757B}.\\

Assuming azimuthal symmetry the magnitude of the magnetic field $B = ({\vec B} .{\vec B})^{1/2}$ is generally expressible as a sum:

\begin{equation}
B(r, \theta) \approx \sum_{l=0}^{l_{max}} B_{l} (r)  Y_l^0 (\theta). 
\end{equation}

It has been seen using simulations~\cite{1968ApJ...153..807H} that the monopole term dominates significantly over the others. We shall ignore the anisotropies as a first approximation in our work.
We emphasize that the vector magnetic field $B$ cannot have a monopole component. The "monopole" here refers to the norm of the magnetic field. We also recognize that stability analysis of the NSs in the presence of strong magnetic fields would naturally require the presence of toroidal and poloidal fields\cite{10.1111/j.1365-2966.2008.14034.x}. In principle, it has been shown that the toroidal fields can be much stronger for realistic stars than the poloidal fields~\cite{10.1111/j.1365-2966.2008.14034.x,2006A&A...450.1077B}.
We have adopted a simplified pure radial profile for the monopolar term of $B$ (it must be noted that we are not looking at magnetic monopoles here) as a polynomial fit function obtained in~\cite{2021EPJC...81..698P,Psaltis_2014}  given by:
\begin{figure*}[htp!]
\begin{tabular}{c}
\includegraphics[width=7.75cm]{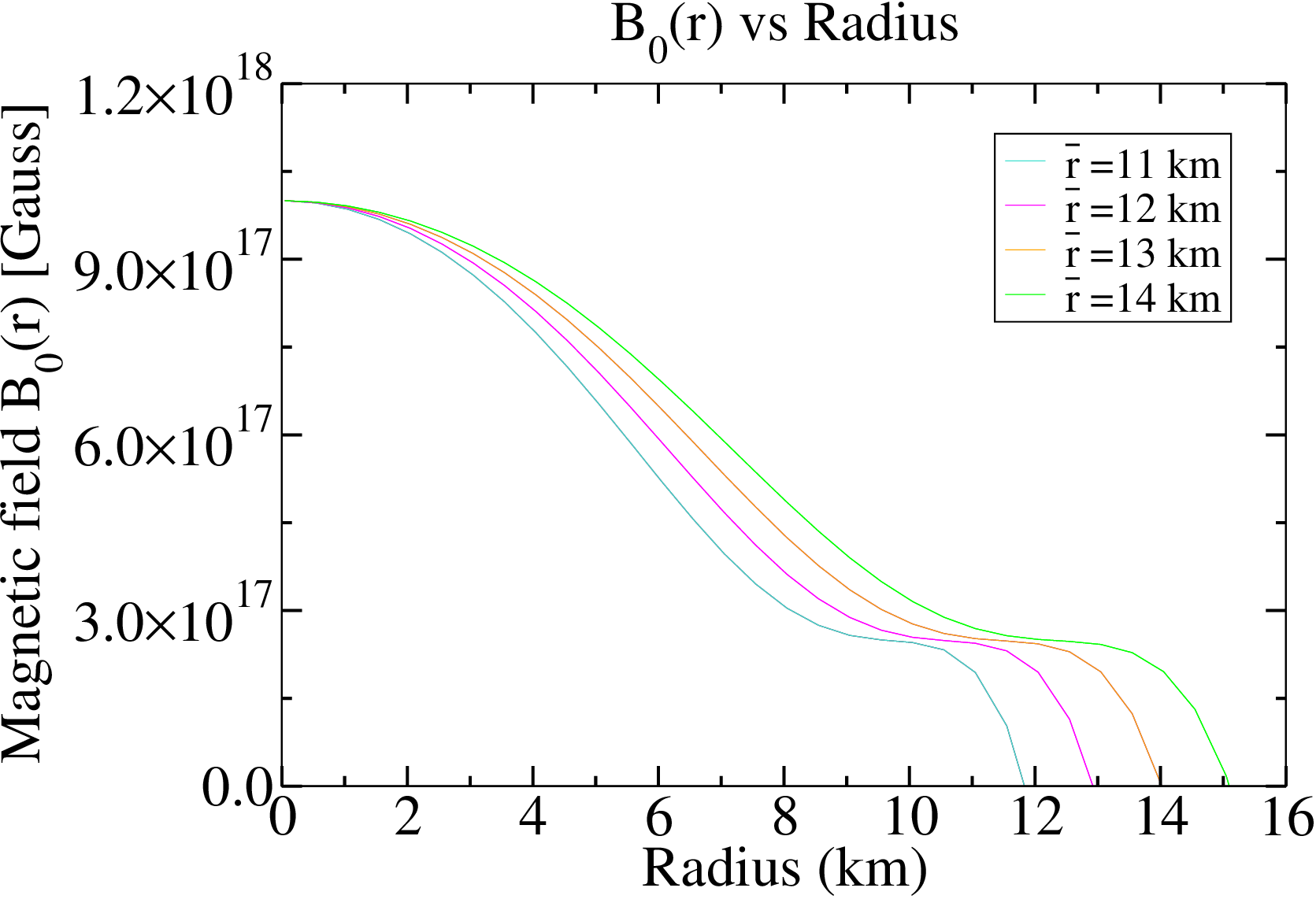}
\end{tabular}
\textbf{\caption{The variation of magnetic field $B_{0}$ (r) with radius for different values of $\overline{r}$}}  
\label{fig:brplot}
\end{figure*}
\begin{equation}
B_0(r) = B_c\left [ 1 - 1.6 \left(\frac{r}{\bar r} \right )^2  -  \left(\frac{r}{\bar r} \right )^4 + 4.2  \left(\frac{r}{\bar r} \right )^6 -2.4  \left(\frac{r}{\bar r} \right )^8 \right],
\end{equation}
where $\bar r$ is the star's mean radius and $B_c$ is central magnetic field. Finding a suitable mean radius of a star distorted due to a magnetic field is difficult. Therefore,  the  $\bar{r}$ value is usually considered to be slightly larger than the star's actual radius. We have assumed the value of central magnetic field, $B_c = 10^{18}$ Gauss, and adopted a value $\bar r$ = 14 km in our analysis.   
One can expect such magnetic fields to decay in relatively short times~\cite{1995A&A...301..757B}. We emphasize that the time evolution of the magnetic field is not considered in the current work, and we have considered only a time-averaged magnetic field with spatial variation. Our work follows the time-independent analysis by ~\cite{2015SSRv..191..239P}.
Figure (\ref{fig:brplot}) shows the radial profile of the magnetic field for different values of the star's mean radius. Our subsequent analysis indicates that $\bar{r}$ has a  weak impact on the luminosities. \\

\subsection{Equation of State}
\label{eos}
In the current work, APR, FPS, and SLY EoSs are considered to study the thermodynamic behavior of the NSs interiors. In Akmal-Pandharipande-Ravenhall (APR) \cite{Gusakov_2005,PhysRevC.58.1804}, EoS interaction potential is parameterized using baryon density and isospin asymmetry~\cite{Haensel_2002,Schneider_2019}. This EoS shows the transition between the low-density phase (LDP) to the high-density phase (HDP). It is also observed that phase transition speeds up the star's shrink rate. The FPS EoS gives a unified description of the inner crust and the liquid core. In this model, the crust-liquid core transition takes place at higher density $\rho_ {edge}$ = $1.6 \times 10^{14}$ gm/cm$^{3}$ and is preceded by a sequence of phase transitions between various nuclear shapes. All phase transitions are weakly first-order, with relative density jumps smaller than $1\%$. In the case of a different effective nuclear Hamiltonian, FPS EoS is the most suitable~\cite{Flowers:1976ux}. The SLY EoS ~\cite{Broderick_2000} of NSs matter is based on the SLY effective nuclear interaction model. In this model, nuclei in the ground state of NSs matter remain spherical down to the bottom edge of the inner crust. Transition to the uniform ionized matter (npe) plasma occurs at $\rho=1.3\times 10^{14}$ g/cm$^{3}$. The SLY EoS consistently describes the cooling near the minimum mass of the solid crust and liquid core. The crust-core phase transition is very weakly first-order, with a relative density jump of about $1$\%. It is mandatory to use the same effective nuclear Hamiltonian to describe nuclear structures (nuclei) and neutron gas inside the inner crust and the uniform npe matter of the liquid core. In the vicinity of the crust-core interface, the SLY EoS is stiffer than the FPS one. Moreover, in the case of the SLY EoS, the discontinuous stiffening at the crust-core transition is more pronounced than in the FPS case.

\subsection{Neutron star cooling}
\label{nsc}
Neutrino emission is the dominant process that causes NSs cooling~\cite{wijnands2017cooling,PRAKASH1994297,PhysRevD.51.348,PhysRevLett.44.1637} at early times and high temperatures. The NSCool is a FORTRAN-based, NSs cooling computational program~\cite {2016ascl.soft09009P}. This package also includes several pre-built stars, many EoSs to build stars, and a TOV integrator to build stars from an EoS. The algorithm solves the heat transport and energy balance equations. 
Assuming the star interior is isothermal, and by focusing only on the energy conservation equation, one can have a reasonable understanding of the fundamental characteristics of NSs cooling~\cite{Beznogov_2023,Yakovlev_2005, PhysRevD.37.2042, PAGE2006497, YAKOVLEV2004523, PhysRevLett.106.081101,PhysRevLett.120.182701,Heinke_2009}. The corresponding equation following Newtonian formulation is as follows~\cite{Buschmann:2021juv}:
\begin{figure*}[htp!]
\begin{tabular}{cc}
\includegraphics[width=7.75cm]{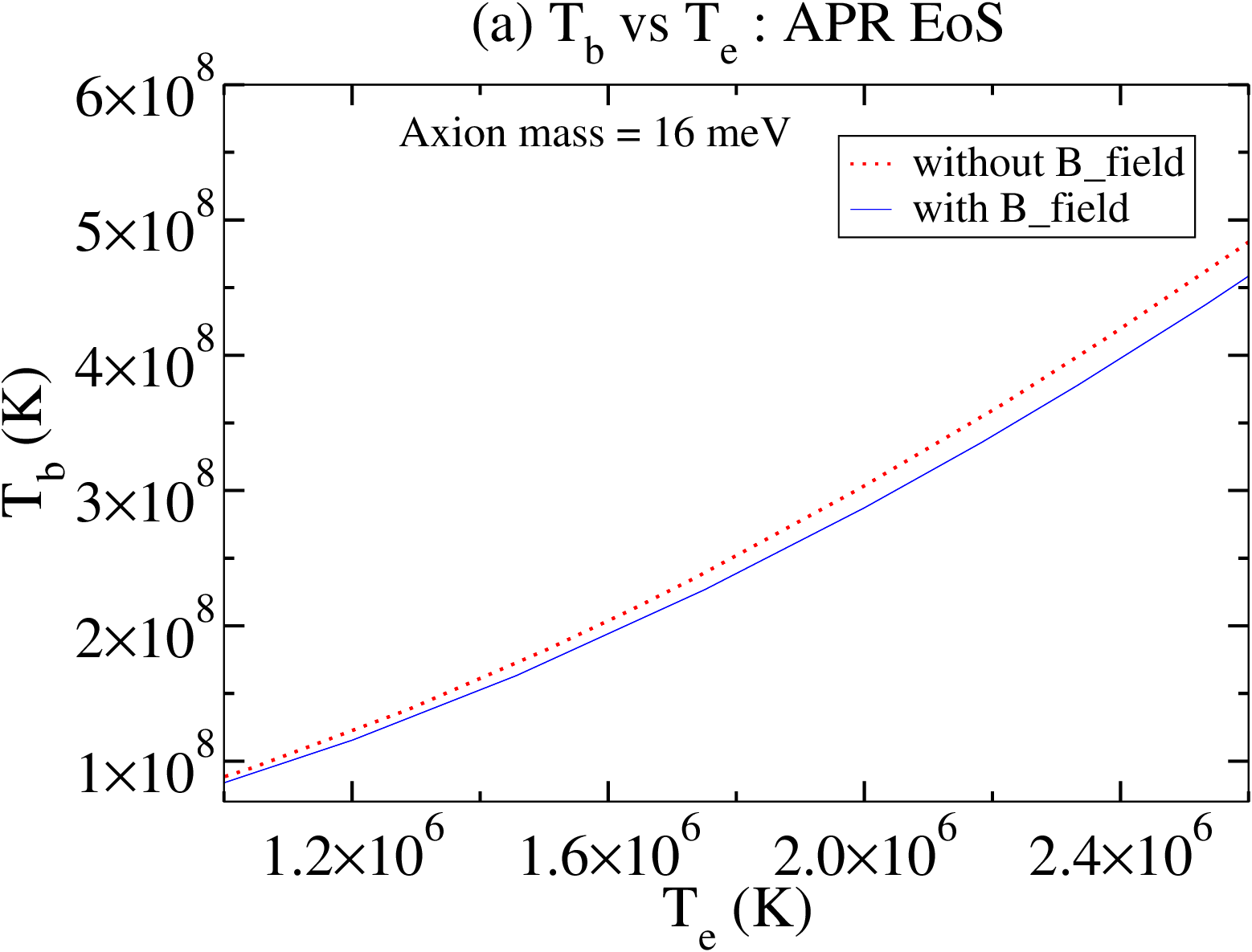}\ & \includegraphics[width=7.75cm]{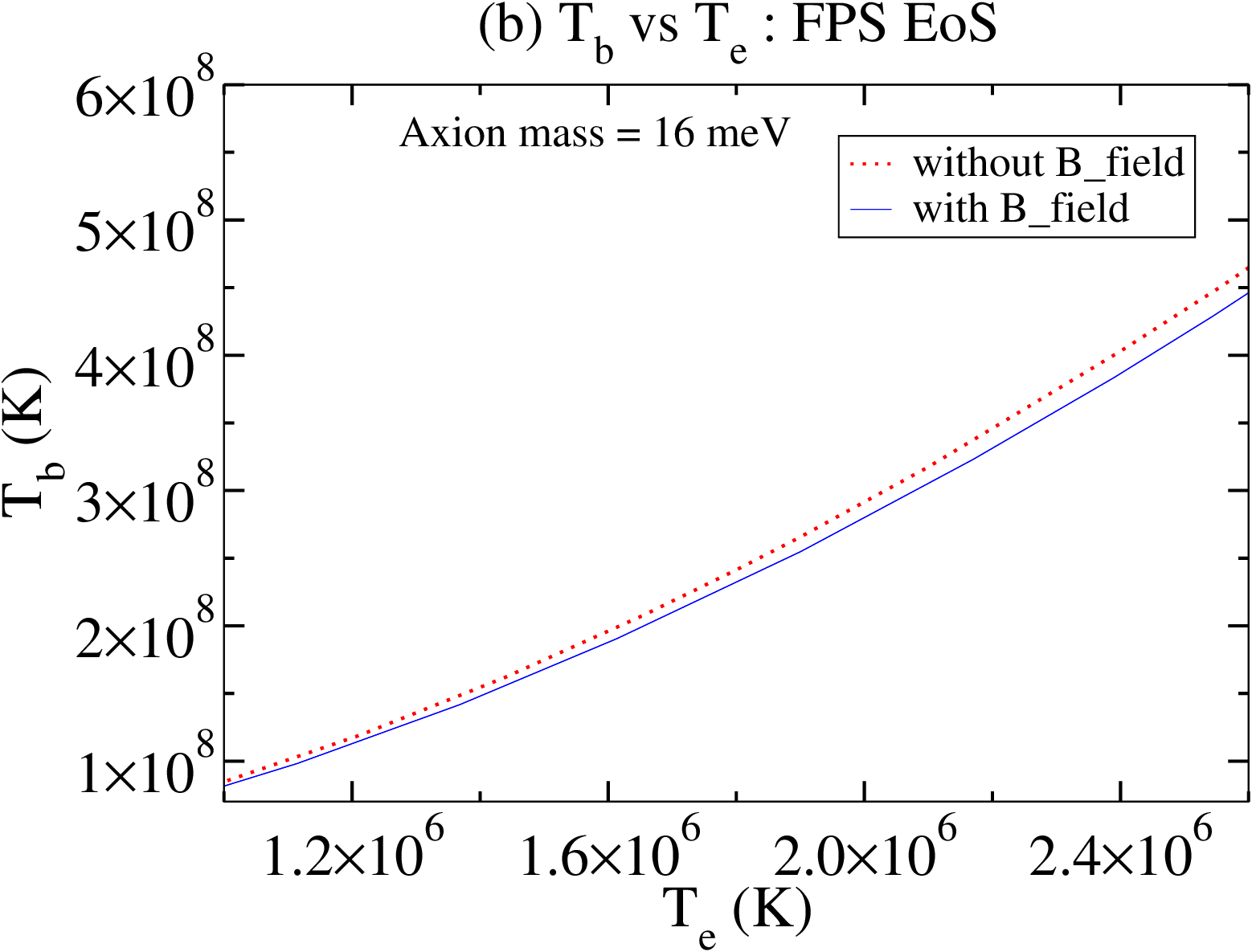} 
\end{tabular}
\end{figure*}
\begin{figure*}[htp!]
\begin{center}
\includegraphics[width=7.75cm]{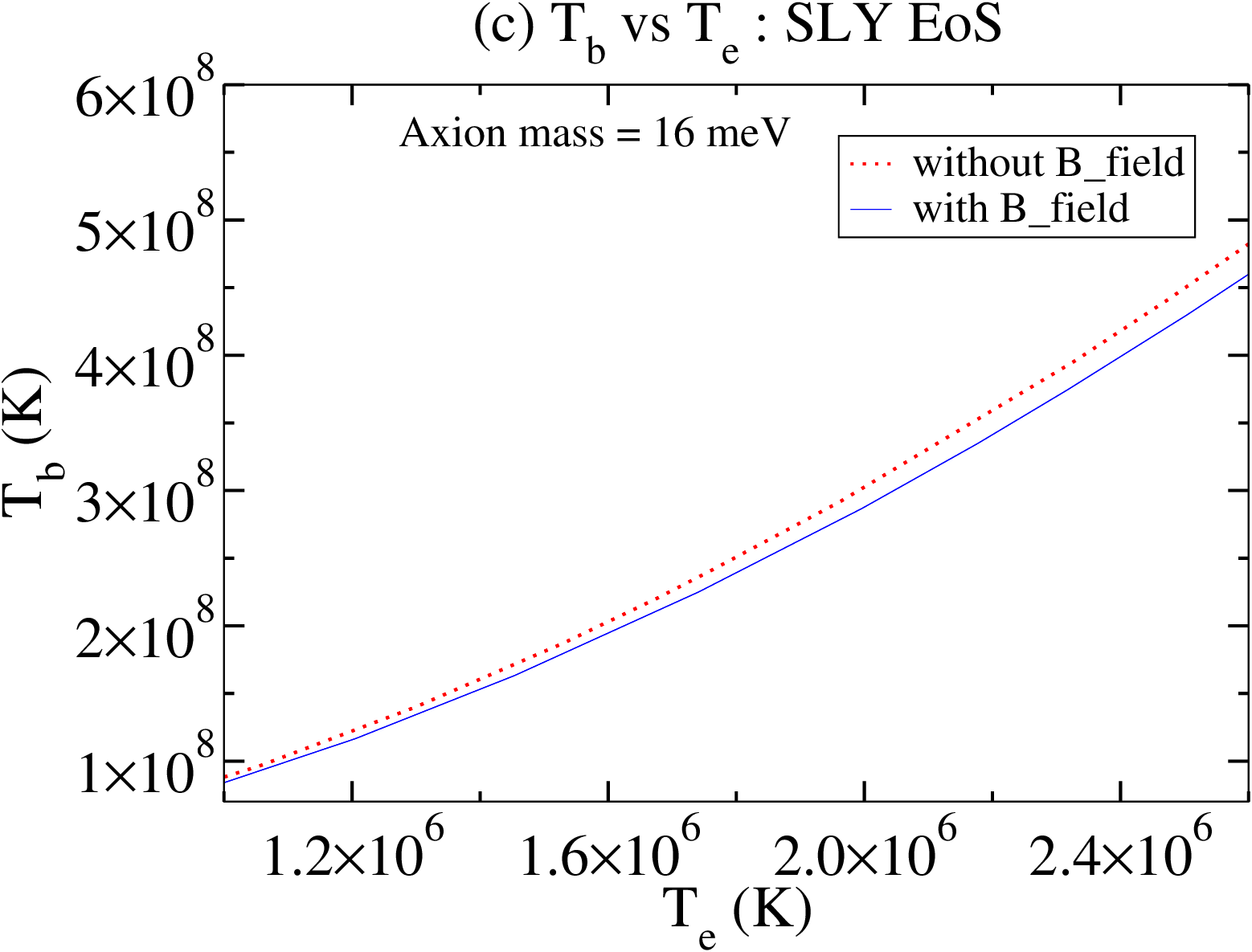}
\end{center}
\caption{The variation of effective surface temperature $T_e$ with temperature $T_b$ for three different EoS in the presence and absence of magnetic field}  
\label{fig:TeTb}
\end{figure*}

\begin{equation}
C_{v}\frac{\mathrm{dT_b^{\infty}} }{\mathrm{d} t}=-L_{\nu }^\infty-L_{a}^\infty-L_{\gamma }^\infty(T_{s})+H,
\end{equation}
where
$C_{v}$ total specific heat,
$L_{\nu }^\infty$ is the energy sinks are the total neutrino luminosity, 
$L_{a}^\infty$ is the axion luminosity, 
$L_{\gamma }^\infty$ is surface photon luminosity, 
$H$ includes all possible heating mechanisms,
$T_{s}$ is the surface temperature and 
$T_b^{\infty}$ internal temperature.
$L_{\gamma }^\infty$= $4\pi\sigma R^{2}(T_{s}^{\infty})^4$, with $\sigma$ as Stefan-Boltzmann's constant and $R$ is the radius of the star.  $T_{s}^{\infty}$ = $T_{s}\sqrt{1-2GM/c^{2}R}$, Here infinity superscript indicates that the external observer is at infinity and measures these quantities on the redshifted scale. Typically, $T_{s}^{\infty}/T_{s}$ $~\sim$ $0.7$.
The cooling profile in the inner layers of the NSs obtained from Eq.(12) gives a simplistic isothermal model of thermal evolution. Generalizations involving multidimensional heat transport equation \cite{2015SSRv..191..239P} may provide more realistic results for NS ages $t < 100 {\rm yrs}$.
 One may obtain approximate scaling laws for the cooling process whereby the effective temperature $T_e$ scales with time as $T_e \sim t ^\nu$.
The exponent $ \nu = - 1/12$ and $\nu = -1/8$ for slow and fast neutrino cooling, respectively. Here we assume $H=0$ and superfluidity in the entire NSs cooling.
Assuming the thermal relaxation time scale of the envelope is shorter than the stellar evolution time scale, and there is minimal neutrino emission within the envelope. In that case, heat transport and hydrostatic equilibrium can be reduced to ordinary differential equations. These equations can be easily solved, provided they have the correct physical input. The surface temperature $T_{s}\equiv T_{e}$ is the outcome for every given $T_{b}\equiv T\left ( \rho _{b} \right )$. Typically, it's referred to as a $T_{b}-T_{s}$ or $T_{b}-T_{e}$ relationship.
The fit functions for the $T_{b}-T_{e}$ relation are given in the literature in the absence of magnetic field. We have calculated the  $T_b-T_e$ relationship  using NSCool code~\cite {2016ascl.soft09009P}. The variation of the observed temperature $T_e$ with the internal temperature $T_b$ is shown in Figure (\ref{fig:TeTb}). The $T_b-T_e$ curves show that at a fixed value $T_b$, the surface temperature $T_e$  in the presence of a strong magnetic field is higher than $T_e$ in the absence of the magnetic field \cite{Potekhin_2003}. The monotonic rise of $T_e$ with $T_b$ is similar to the behavior reported earlier by 
\cite{Buschmann:2021juv}.The $T_b-T_e$ relationship has been estimated in many previous works \cite{1983ApJ...272..286G,1987ApJ...312..711N,1997A&A...323..415P}.
Considering a very intense magnetic field in hot blanketing envelopes of NSs, the structure of HBEs may be affected by enhanced neutrino emission \cite{refId01,potekhin2007heat,refId02,Potekhin_2003,2016MNRAS.459.1569B,10.1093/mnras/stu1102}. The boundary of the heat blanketing envelopes (HBEs) is $\epsilon_b$ ($\rho_{b} \sim 10^{10}$ gm/cm$^3$). The magnitude of the magnetic field at HBEs is taken as $10^{13}$ Gauss and it has a radial dependance given by Eq.$(10)$. The  HBEs contain either iron or successive layers of hydrogen, helium, carbon and iron. 
The radiative opacity for a fully ionized, non-relativistic, and non-degenerate plasma, composed of electrons and ions is determined by the Kramers' formula given by: \cite{BEZNOGOV20211}: 
\begin{equation}
K_{r}(0)= 75 \bar{g}_{eff}\left (\frac{Z^{3}}{A^{2}} \right )\rho \left ( \frac{10^{6}K}{T} \right )^{3.5},
\end{equation}
where $A$ and $Z$ are relative atomic weight and mass number, respectively, and $\bar{g}_{eff} \sim 1$ is the effective Gaunt factor, $\rho$  is measured in gm/cm$^3$. 
The strong magnetic fields affects the thermal conduction in the outer layers of the NS, thereby affecting heat flows near the stellar surface and the effective surface temperature. The relation between radiative opacities arising from free-free transitions  is given by~\cite{BEZNOGOV20211}:
\begin{equation}
K_{r}(B)=2.2\bar{g}_{eff}Z^{3} \rho~ A^{-2}\left ( \frac{10^{6}K}{T} \right )^{1.5}\left ( \frac{10^{12} G}{B} \right )^2,
\end{equation}
where $\bar{g}_{eff} \sim 1$ is the effective Gaunt factor and $\rho$  is measured in gm/cm$^3$.
We note that though the radiative conductivity becomes anisotropic in the presence of strong magnetic fields.

Furthermore, a magnetic field inside the heat blanketting layer of NSs can strongly affects the electron thermal conductivity. Here we consider only thermal conductivity of electrons which becomes a tensor quantity (wth three non-zero components) in the presence of a magnetic field. The three thermal conductivities coefficients are given by parallel ($\kappa_{\parallel}$) and transverse $\kappa_{\perp}$ to the field and hall $\kappa_{H}$ which describes component perpendicular to B and temperature gradient. The effective electron thermal condctivitiy is given by~\cite{10.1111/j.1365-2966.2007.12301.x,refId01,BEZNOGOV20211}:
\begin{equation}
\kappa =\kappa_{\parallel} \cos^{2}\theta_B +\kappa_{\perp}\sin^{2}\theta_B,
\end{equation}
where
\begin{equation}
\kappa _{\parallel }=\frac{\pi ^{2}k_{B}^{2}T\,n_{e}\tau_{e}}{3m_{e}^*},
\end{equation}
In the above equation, $\tau_e$ is the effective electron thermal-conduction relaxation time and $n_e$ is the electron number density.  
\begin{equation}
\kappa_{\perp }=\frac{\kappa_{\parallel }}{1+(\omega^{*}\tau_{e})^{2}}, 
\end{equation}
where $\omega^{*} = \omega_c/\gamma_r$, with $\omega_c=e B/m_e c$, $\gamma_r= \sqrt{(1+x_r^2)}$ and $x_r = \frac{p_F}{m_e c}$. Finally, the product $\omega^{*} \tau_e=1760 (\frac{B_{12}}{\gamma_r}) (\frac{\tau_e}{10^{-16} s})$, with $B_{12}=B/10^{12} \mbox{Gauss}$. Here $\theta_B$ is the angle between the direction of magnetic field and normal to the surface.   

\subsection{QCD Axions}
The axions are pseudo-Nambu-Goldstone boson~\cite{Peccei:1977hh,Peccei:1977ur,Weinberg:1977ma,PhysRevLett.40.279,PhysRevLett.123.021801,umeda1998axion} emerging as Dark matter (DM) particles. These are motivated by the Peccei-Quinn solution to the strong-CP problem in which a theta term in QCD Lagrangian is canceled by introducing a scalar field (the so-called QCD Axions)~\cite{ABBOTT1983133,PhysRevD.32.3172,Dessert_2022}. Due to astrophysical and cosmological constraints, axions must have rest mass $m_{a} \sim 10^{-6}$ eV to $10^{-2}$ eV ~\cite{PRESKILL1983127,DINE1983137,Buschmann:2019icd,chadha2022axion,adams2022axion}.
In order to study the production of DM Axions, we need NS cooling~\cite{Buschmann:2019pfp,Duffy_2009}. There is continuous heat transport from the interior of NS to the surface. 
These light particles~\cite {Marsh_2017} may be emitted from the interiors or the surface of the NSs. In fact, axions in an extended NSs magnetospheres~\cite{PhysRevLett.123.061104} can couple to virtual photons and produce the real photons due to Primakoff effect ~\cite{PhysRevD.92.075012,Pshirkov_2009}.
As per the current status of the Axionic Dark Matter experiment (ADMX), it can be concluded that axions may remain an excellent candidate for DM searches~\cite{PhysRevLett.120.151301,PhysRevLett.117.141801,PhysRevD.97.123006,PhysRevLett.117.141801,Marsh_2017}.
There are two axion models; the Dean-Fischler-Srednitsky-Zhitnitsky - DFSZ model~\cite{Sedrakian_2016,Leinson_2019}, which includes an additional axion coupling to the charged leptons, while the Kim-Shifman-Weinstein-Zakharov (hadronic) - KSVZ model restricts the Axion's interactions to photons and hadrons. 
\subsection{ Neutrino-Axion emission inside NS core}
\label{aer}
\subsubsection{Cooper Pair Breaking and Formation process (PBF)}
\label{cooper}
Microscopic theories of dense matter play a crucial role in cooling compact stars and investigating the composition of the inner core. Both the neutrino and photon emissions are controlled by the supranuclear densities and the structure of the stellar outer layers~\cite{Page:2005fq,Buschmann:2019pfp,Page_2009,Leinson_2000}, respectively. 
The low value of Fermi momenta, particularly for nucleon pairing, is predicted to occur
in the $^1S_0$ angular momentum state. Cooper pairs are expected to form and break when the superfluid condensate is in thermal equilibrium with the broken pairs. This usually happens at a temperature far below superfluid critical temperature $T_{c}$. Neutrino and Axion can be produced due to the formation of a Cooper pair liberating energy which can be taken away by a $\nu\bar{\nu}$ pairs.

\begin{align} 
n+n\to \left [ nn \right ] + \nu +\overline{\nu}\nonumber\\
p+p\to \left [ pp \right ] + \nu +\overline{\nu}\nonumber
\end{align}
 
In the current work, we have not included the neutrino emission processes namely, the direct and modified Urca processes.

\subsubsection*{Neutrino emission rate} 
\label{ner}

The substantial cooling of NSs with superfluid inside the core by emitting neutrinos through Cooper pair-breaking processes (PBF)~\cite{Buschmann:2019pfp,Sedrakian_2016,Keller_2013,Page:2005fq,Page_2009} is driven by axial-vector currents when the temperature~\cite{Geppert_2006,Kolomeitsev_2008,PhysRevLett.66.2701} lies below the critical temperature $T_{c}=10^{9}$ K. For neutron/proton $^1S_{0}$-wave paired superfluid, the emissivity~\cite{2001PhR...354....1Y,Keller_2013,Leinson_2000} of neutrino is given by:

\begin{equation}
\epsilon_{\nu }^s = \frac{5\,G_{F}^2}{14\pi^{3}}v_{N}\left ( 0 \right )v_{F}\left ( N \right )^2T^7I_{\nu}^s.
\end{equation}
Here the integral $I_{\nu}^s$ is;

\begin{equation}
I_{\nu}^s=z_{N}^7\left ( \int_{1}^{\infty }\frac{y^5}{\sqrt{y^{2}-1}} \left [ f_{F}\left ( z_{N}y \right ) \right ]^2 dy\right ),
\end{equation}
where $\epsilon_{\nu}^s$ is neutrino emissivity and $G_{F}$ is Fermi's coupling constant $=1.166\times 10^{-5} \mbox{GeV}^{-2}$. $z=\Delta(T)/T$ with $\Delta(T)=3.06\,T_c\,\sqrt{(1-\frac{T}{T_c})}$. Here $T_c $ is the critical temperature for neutron/proton superfluid. 

The neutrino emissivity due to the P-wave paired neutron superfluid is taken from ref.~\cite{Page:2005fq}.\\
\subsubsection*{Axion emission rate}
In this section, we briefly describe the axion emission rate inside NSs core~\cite{PhysRevLett.123.061104,PhysRevLett.53.1198} from Cooper pair breaking and formation process (PBF)~\cite{Sedrakian_2016} process. In order to calculate the production rates in the NSs core, prerequisites are the temperature profiles in the core, the metric, the critical temperature profiles, neutron and proton Fermi's momenta profile (which depends on the equation of state). The modified TOV system of equations (in the presence of magnetic field and their counterparts) profiles are used after making specific changes (to include axion emission) in the NSCool code to determine the cooling rate and luminosities. 
Both spin-$0$ S-wave and spin $1$ P-wave nucleon superfluids could exist inside the NS core. The axion emission rate~\cite{Buschmann:2019pfp,Sedrakian_2016} due to the neutron S-wave pairing from the Cooper pair-breaking formation (PBF)~\cite{Keller_2013} is given by:

\begin{equation}
\epsilon_{a}^s= \frac{8}{3\pi f_{a}^{2}}\,v_{n}(0)\,v_F(n)^{2}\,T^{5}\,I_{a}^s.
\end{equation}
The integral $I_{a}^s$ is expressed as:

\begin{equation}
I_{a}^s= z^5_{n}\left ( \int_{1}^{\infty} \frac{y^3}{\sqrt{y^2-1}}\left [ f_{F}\left ( z_{n}y \right ) \right ]^{2} dy\right),
\end{equation}
where
$\epsilon_{a}^s$ is axion emissivity,
$f_{a}$ is axion decay constant,
$v_{n}(0)$ density of state at Fermi surface and
$v_F(n)$ is fermi velocity of neutron.

\begin{eqnarray}
v_{n}\left ( 0 \right )=\frac{m_{n}\,p_{F}\left ( n \right )} {\pi^2} \\
z=\frac{\Delta(T)}{T}, \\
f_{F}\left ( x \right )= \left [ e^x+1 \right ]^{-1},
\end{eqnarray}
with $x=\frac{\omega}{2T}$, where $\omega$ is the axion energy. \\

The ratio of axion and neutrino emissivities is given by: 
\begin{equation}
{\epsilon _{a}^s} = \left ( \frac{59.2}{f_{a}^2G_{F}^2[\Delta\left ( T \right)]^2}r(z) \right )\epsilon_{\nu}^s,
\end{equation}
where, 
\begin{align}
\Delta(T)\simeq 3.06\,T_{cn}\sqrt{1-\frac{T}{T_{cn}}}.
\end{align}
The numerical values of the $r(z)=z^2\,I_a^s/I_{\nu}^s$ associated with axion and neutrino emissivity integrals for the different values of $z$ around one are always less than unity.

\begin{equation}
f_{a}> 5.92\times 10^9 GeV\left [ \frac{0.1 MeV}{\Delta\left (T\right)} \right ].
\end{equation}
Here $f_{a}$ is a axion model-dependent variable ~\cite{Buschmann:2021juv}. Axion mass $m_{a}$ is related to $f_{a}$ by the equation;~\cite{Sedrakian_2016,Leinson_2019,PhysRevD.34.843,PhysRevD.37.1237}:

\begin{equation}
m_{a}= 0.60\; \text{eV}\times \frac{10^{7}\text{GeV}}{f_{a}}.
\end{equation}
Here, we have assumed the axion emissivity due to proton S-wave pairing~\cite{Sedrakian_2016,1999A&A...343..650Y,1999A&A...345L..14K}  will remain the same as axion due to neutron S-wave pairing provided $C_n=C_p$ and $T_{cn}=T_{cp}$, where $T_{cn}$ and $T_{cp}$ are crtical temperatures for neutron superfluid and proton superfluid, respectively. Here we have taken $T_{cp}=T_{cn}=10^9$ K.\\

 The axion emissivity due to the P-wave paired neutron superfluid is given by~\cite{Sedrakian_2016}:\\
\begin{equation}
\epsilon_{a}^{P}=\frac{2\,C_{n}^{2}} {3\pi f_{a}^{2}}\,v_{n}(0)\,T^{5}\,I_{a}^p,
\end{equation}
where $C_n$ is the axion model dependent constant. 
 
The integral $I_a^p$ is expressed as:
\begin{equation}
I_{a}^{p}=z_{n}^{5}\left ( \int_{1}^{\propto }\frac{y^{3}}{\sqrt{y^{2}-1}} [f_{F}\, (z_n y)]^2 dy\right ),
\end{equation}
where $z_{n}=\Delta^P(T,\theta)/T$ with $\theta$ is the polar angle.  
There exist two states of P-wave superfluid pairing denoted as A and B expressed as: 
\begin{equation}
\Delta_{P}^{A}=\Delta _{0}^{A}\sqrt{1+3\,\cos^{2}\theta}.
\end{equation}
and 
\begin{equation}
\Delta_{P}^{B}=\Delta _{0}^{B}\,\sin\theta.
\end{equation}
\subsection{Neutrino-Axion emission inside NS crust}
\label{nnbrem}
\subsubsection{e-e Bremsstrahlung process}
\label{cooper}
\subsubsection*{Neutrino emission rate}
The neutrino emission rate for e-e Bremsstrahlung process is given by~\cite{PhysRevLett.53.1198,Sedrakian_2016}: 
\begin{equation}
\epsilon_{\nu ee,brem}=7.42 \times 10^{-2}\,G_{F}^{2} \,Z^{2}\,\alpha_e^{4} \,n_{i}T^{6}~L.
\end{equation}
Here $n_{i}$ is nuclei number density, $\alpha_e=1/137$ is the QED fine structure constant, and $L$ incorporates many-body corrections to process rate related to nuclei correlations.

\subsubsection*{Axion emission rate}
The NSs crust may emit axions when a degenerate relativistic electron strikes an ion with charge $Z$ and mass number $A$. The axion emission rate is given by~\cite{PhysRevLett.53.1198,Sedrakian_2016}:
\begin{equation}
\epsilon_{aee,brem}=\frac{\pi ^{2}}{120}\frac{Z^{2}\alpha_e}{A}\left (\frac{g_{aee}}{\epsilon _{F}}\right)^{2}n_{B} T^{4}\left[2\ln(2\gamma)-\ln\frac{\alpha_e}{\pi} \right ].
\end{equation}
Here  $g_{aee}$ is coupling of axions to electrons, $g_{aee}$=$\frac{C_{e}\,m_{e}}{f_{a}}$, $\gamma$ is the Lorentz factor of electrons,  $\epsilon_{F}$ is the electron's fermi energy, $n_{B}$ is baryon number density, and  $\alpha_e=1/137$ is the QED fine structure constant.
By making electron PQ charge $C_{e}$=$1$, L=$1$ and $\frac{m_{e}}{\epsilon_{F}}$= $10^{-2}$ the relation between Axion to neutrino emissivity is given by:
\begin{equation}
\epsilon_{aee,brem} = \left [ 2.8\left ( \frac{10^{9}K}{T} \right )^{2}\left ( \frac{10^{10}GeV}{f_{a}} \right )^{2}  \right ]\epsilon_{\nu ee,brem}.
\end{equation}
We want to emphasize here that we have not considered the influence of strong fields on the mentioned neutrino and axion cooling mechanisms. 

\section{Results and Discussions}
\label{result}

\begin{figure*}[htp!]
\begin{center}
\includegraphics[width=7.75cm]{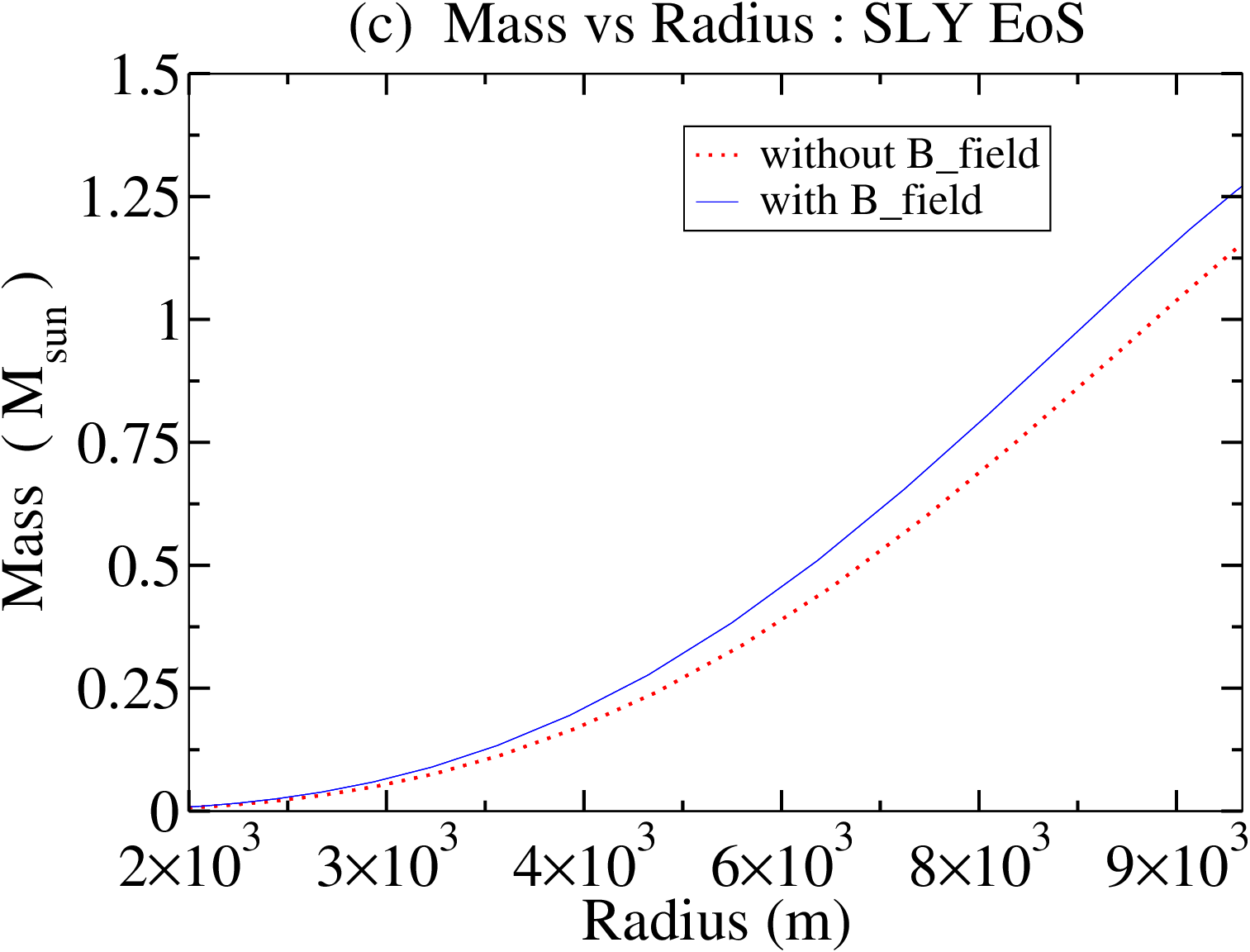}
\end{center}
\caption{The variation of mass with radial distance  $r$ for three different EoSs in the presence and absence of magnetic field.}  
\label{fig:masswithr}
\end{figure*}

Figure (\ref{fig:masswithr}) depicts the variation of mass with respect to the distance from the center of the NS for three different EoSs, namely APR, FPS, and SLY, in the presence of the magnetic field. It indicates that the variation of mass $m(r)$ as a function of distance from the center of the NS change little due to the effect of the magnetic field.  The difference is seen at some intermediate radial distances. The APR, FPS, and SLY models show an approximately $ 16.7\% $,  $15.4\%$ and  $17.5\%$ departure from the corresponding results without including magnetic field  $ r \sim 4500$ m, respectively.\\

\begin{figure*}[htp!]
\begin{tabular}{cc}
\includegraphics[width=7.75cm]{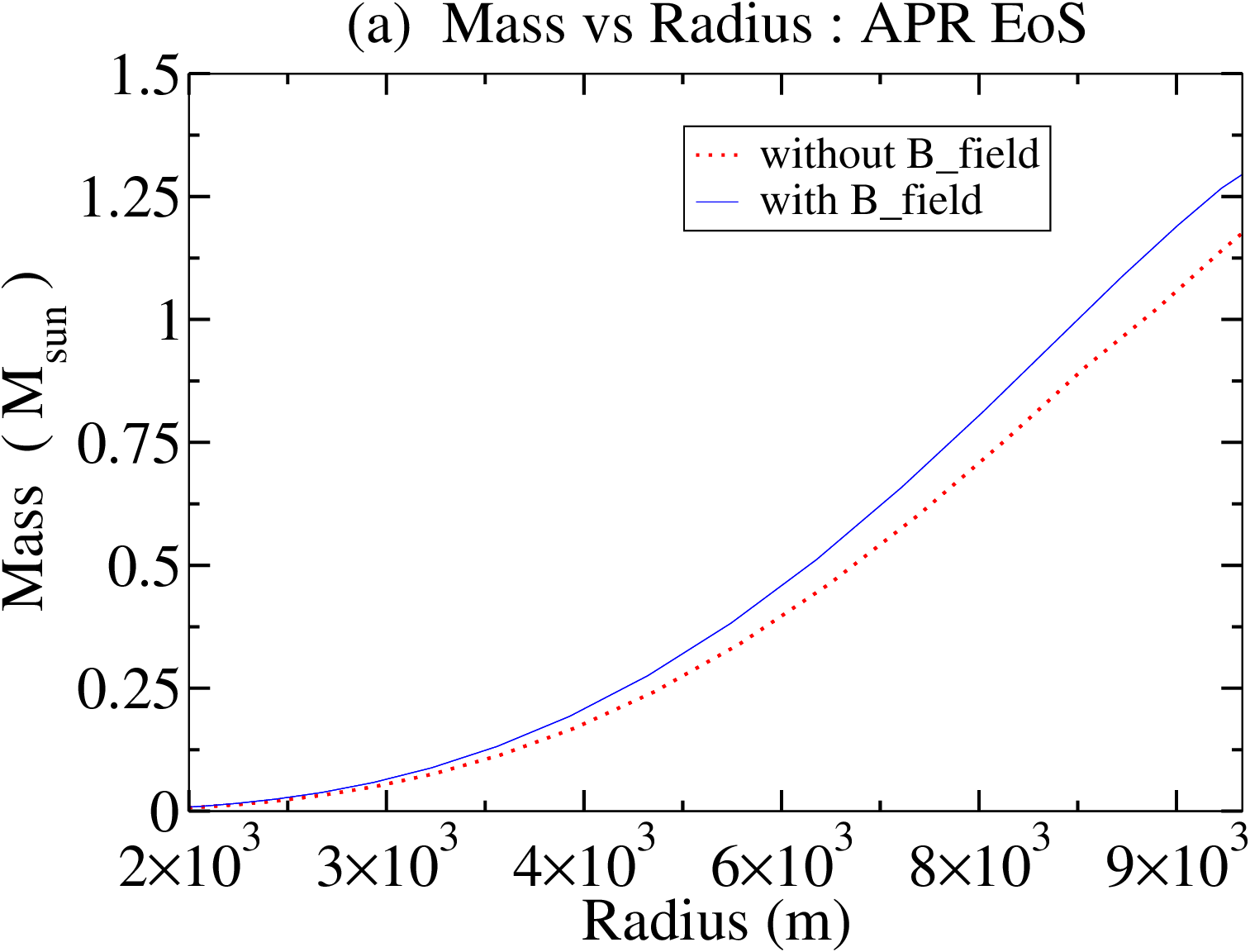}\ & \includegraphics[width=7.75cm]{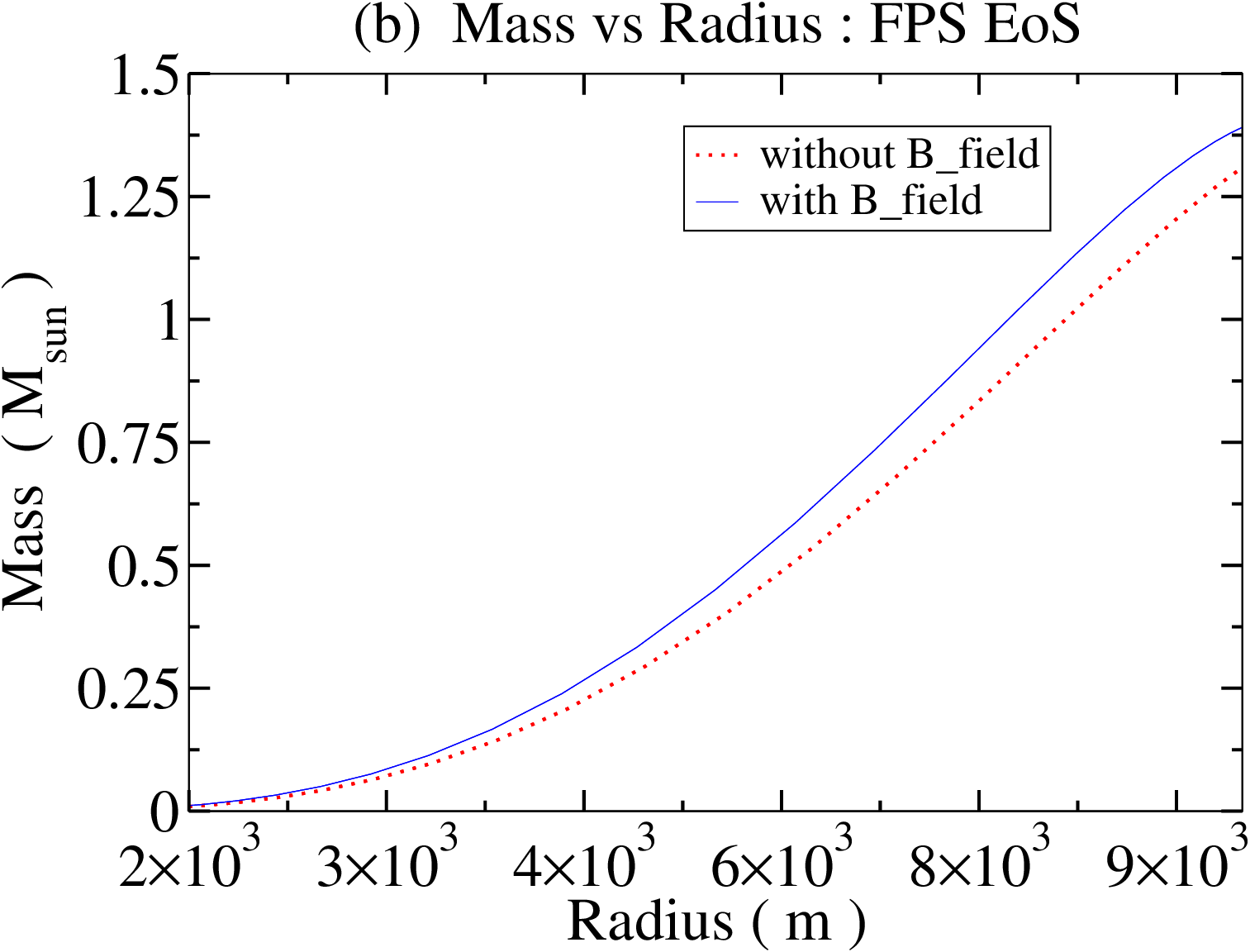} 
\end{tabular}
\end{figure*}

\begin{figure*}[htp!]
\begin{tabular}{cc}
\includegraphics[width=7.75cm]{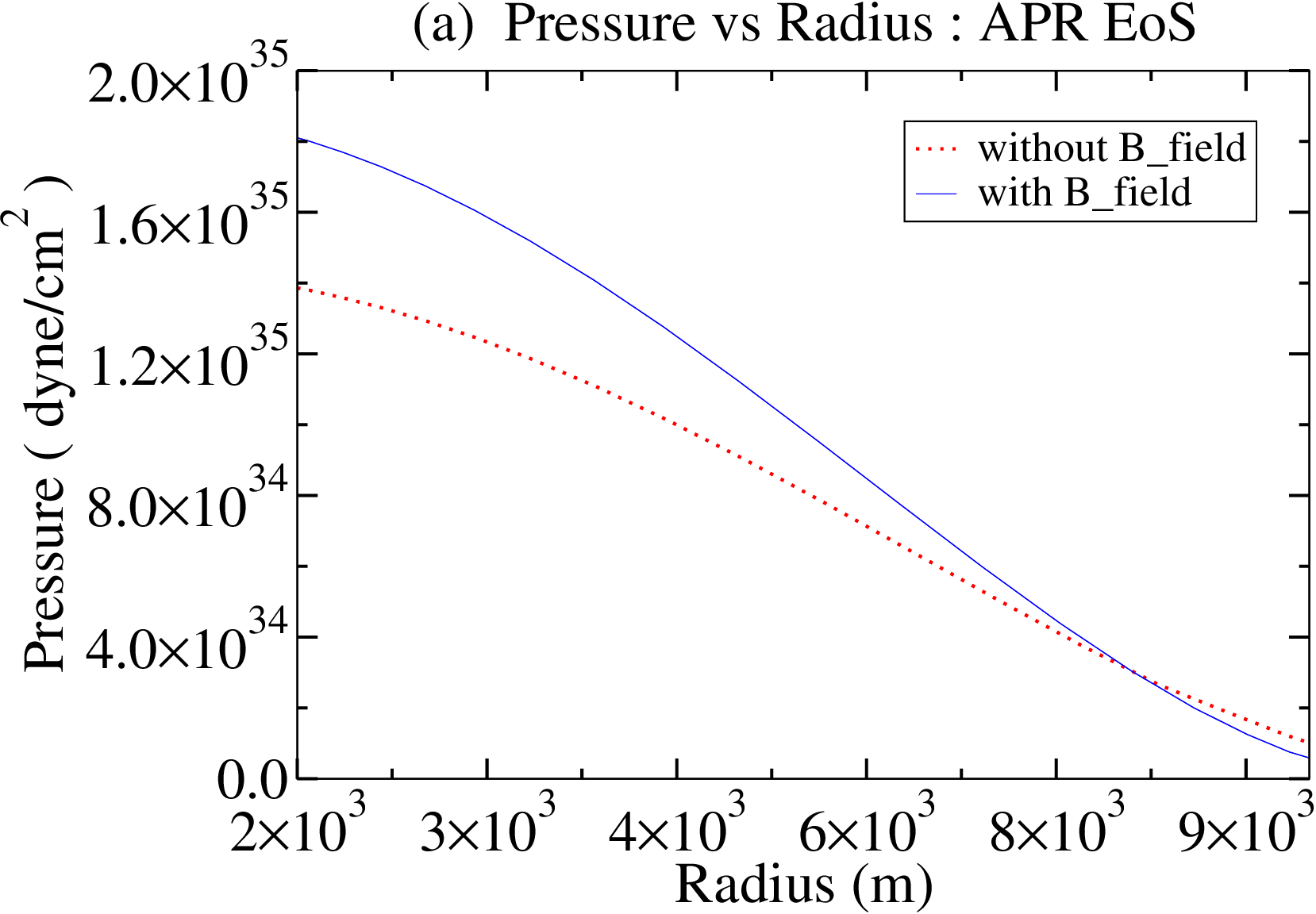}\ & \includegraphics[width=7.75cm]{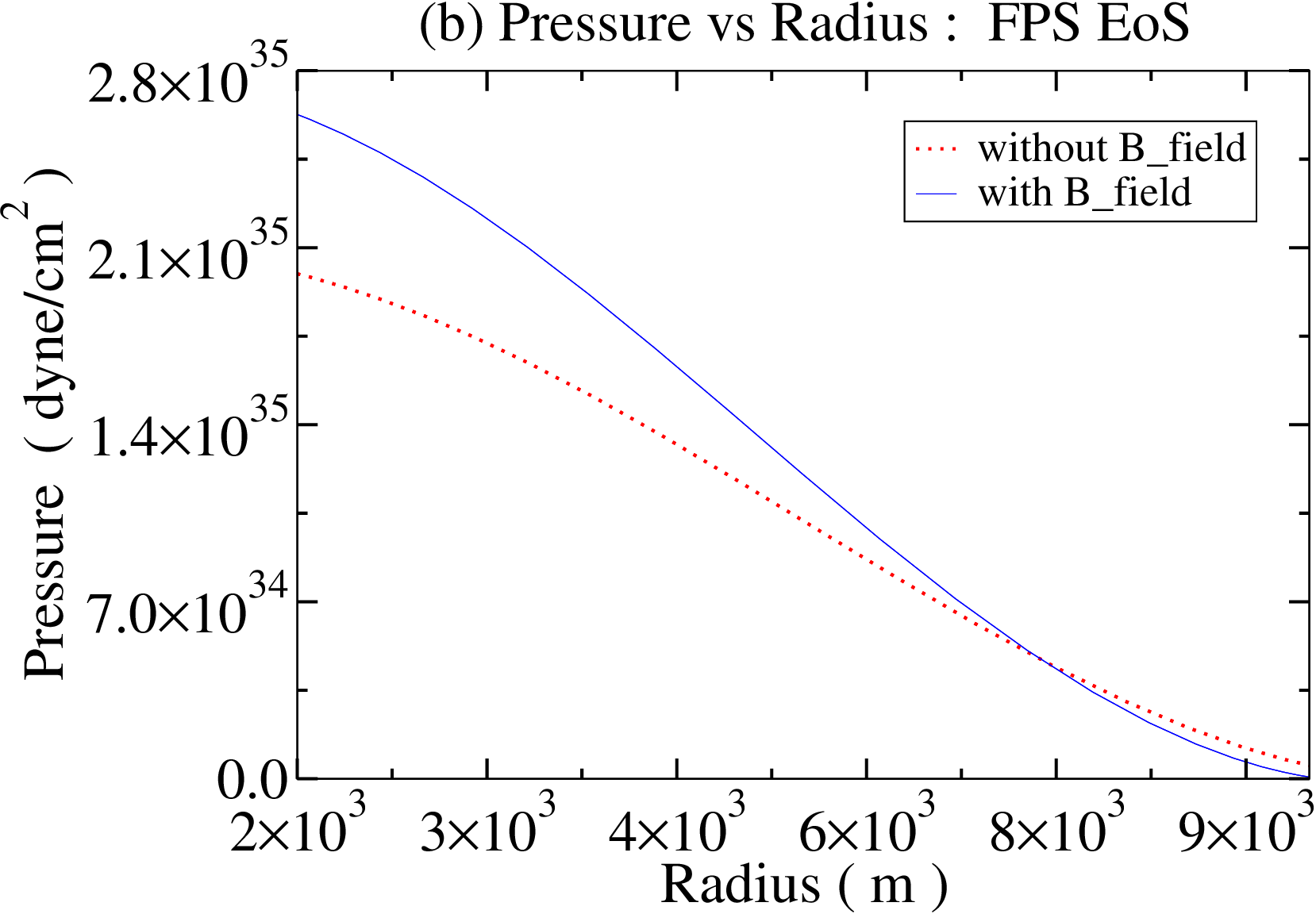} 
\end{tabular}
\end{figure*}
\begin{figure*}[htp!]
\begin{center}
\includegraphics[width=7.75cm]{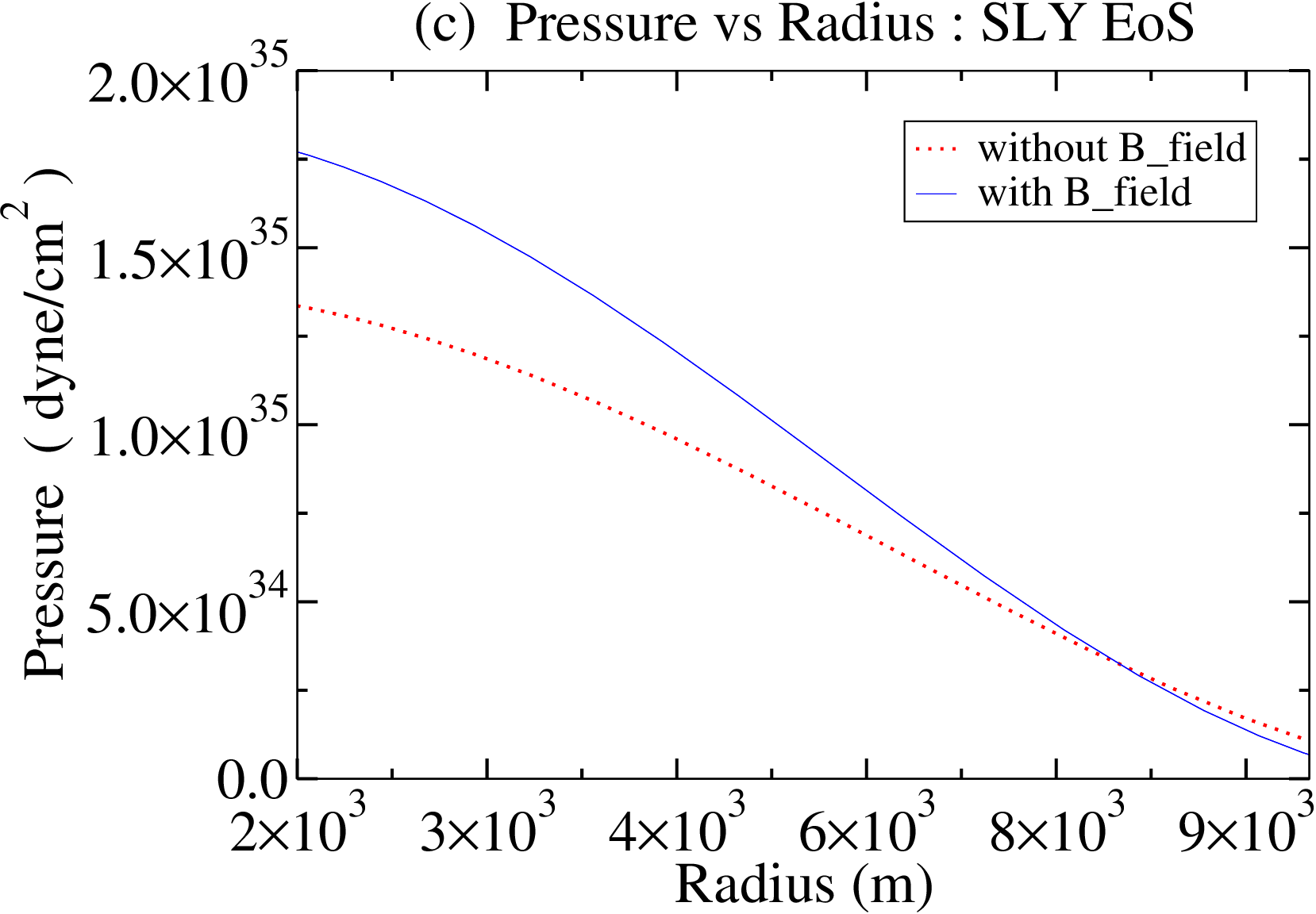}
\end{center}
\caption{The variation of pressure with radial distance  $r$ for three different EoSs in the presence and absence of magnetic field.}   
\label{fig:pressurewithr}
\end{figure*}
Figure (\ref{fig:pressurewithr}) shows the variation of pressure as a function of distance from the center of the NS for three different EoSs, namely APR, FPS, and SLY, before and after taking into account the magnetic field. Here a significant difference due to the magnetic field is seen. The departure from the case with the magnetic field increases with the radial distance from the center. After incorporating the magnetic field, the $P(r)$ curve lies above the corresponding curve without the magnetic field effect and the curves coincide at $ r\sim 8100$ m for APR and SLY model EoSs ,and  $ r\sim 7400$ m for FPS model EoS. On comparing the slope of the two curves, it is found that the pressure decreases slower after including the magnetic field. The behavior for three different EoSs is qualitatively similar, and the APR, FPS, and SLY models show an approximately $ 20.1\% $,  $18.5\%$ and  $20.2\%$ departure with respect to the case of with magnetic field at  $ r\sim 4500 $ m, respectively. The imprint of the magnetic field on the cooling properties of the NSs is studied using the NSCool code~\cite {2016ascl.soft09009P}. \\
\begin{widetext}
\begin{center}
\begin{table}[h]
\label{table:tablenos}
\begin{tabular}{ c | c | c } 
 \hline
 \hline
 $EoS$ \ & $R$ (without B field) (km) \ & $R$ (with B field) (km)  \\ [0.5ex] 
 \hline
  APR \ & $11.77$\ & $10.37$ \\ 

 FPS \ & $10.79$\ & $9.68$ \\

 SLY \ & $11.64$\ & $10.43$\\ [1ex]
 \hline\hline 
\end{tabular}
\caption {  Table showing radius ($R$) of NSs corresponding to the mass  $M = 1.4 M_\odot$  for different EoS.} 
\end{table}
\end{center}
\end{widetext}

\begin{figure*}[htp!]
\begin{center}
\includegraphics[width=7.75cm]{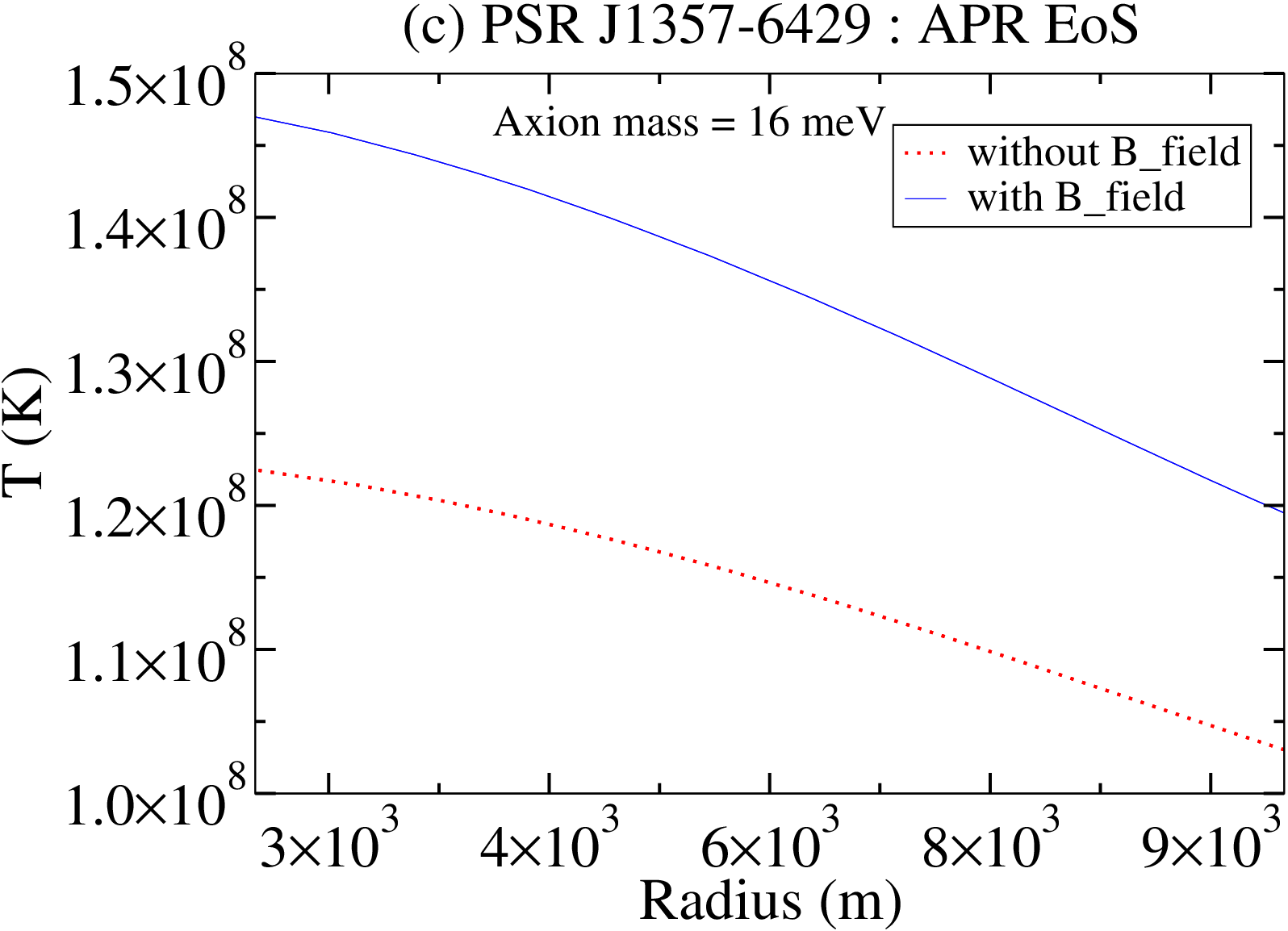}
\end{center} 
\caption{ The variation of temperature with radial distance $r$ for three different NSs for APR EoS in the presence and absence of magnetic field.} 
\label{fig:nsradial1}  
\end{figure*}

In Figure (\ref{fig:nsradial1}), we first look at the temperature profile for axion mass $\sim 16$ meV as a function of radial distance from the center at a particular time corresponding to a characteristic age of a particular NS for the APR EoS. The same is shown for two more NSs having two different characteristic age with axion mass $\sim 16$ meV. The characteristic age used here are $9.7\times 10^{2}$ yrs~\cite{O'Dea_2014, refId0},$3.3\times 10^{4}$ yrs ~\cite{Xu_2021,Ng_2007} and  $7.3 \times 10^{3}$ yrs~\cite{Zavlin_2007,2012ASPC..466...29K} for the three figures in the panel, respectively. From the figures, it is clear that in the presence of a magnetic field, the cooling property of the star undergoes a considerable change. As the star cools, in the absence  of the magnetic field, the temperature remains lesser than the case when there is magnetic field at all distances and at all times.

 For a given star, the presence of a strong magnetic field and an associated high magnetic field energy lead to a decrease of thermal energy while the gravitational energy remains unchanged. This explains the correlation between the suppression of luminosity and the strength of the magnetic field. Similar suppression of temperature in white dwarf stars with magnetic fields is reported in earlier works~\cite{10.1093/mnras/sty776}. The strong magnetic field may affect the EoS and may also cause sizable changes to thermal conduction and observable properties. We have not incorporated the impact of magnetic field on the EoS in the current work.

\begin{figure*}[htp!]
\begin{tabular}{cc}
\includegraphics[width=7.75cm]{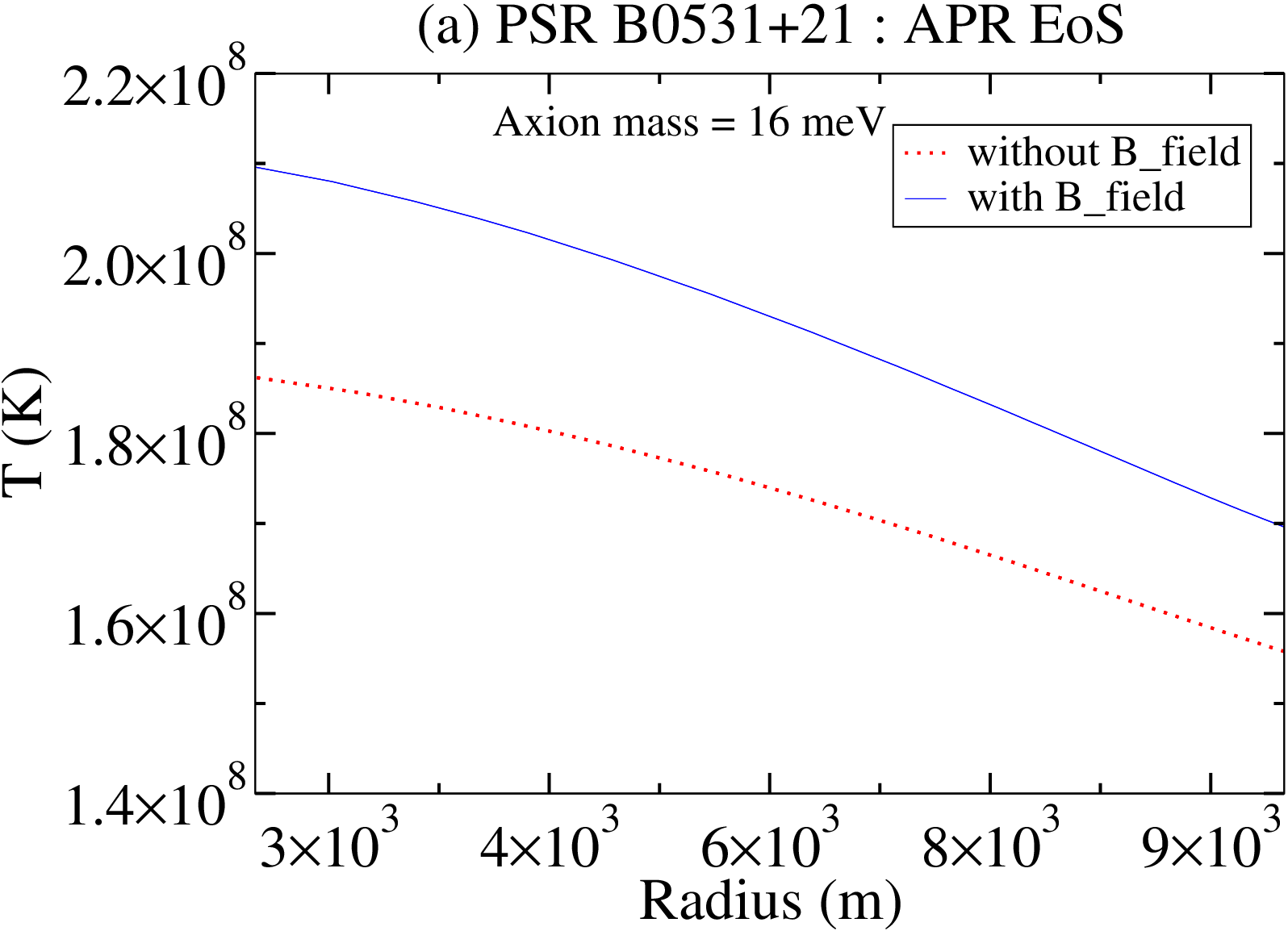}\ & \includegraphics[width=7.75cm]{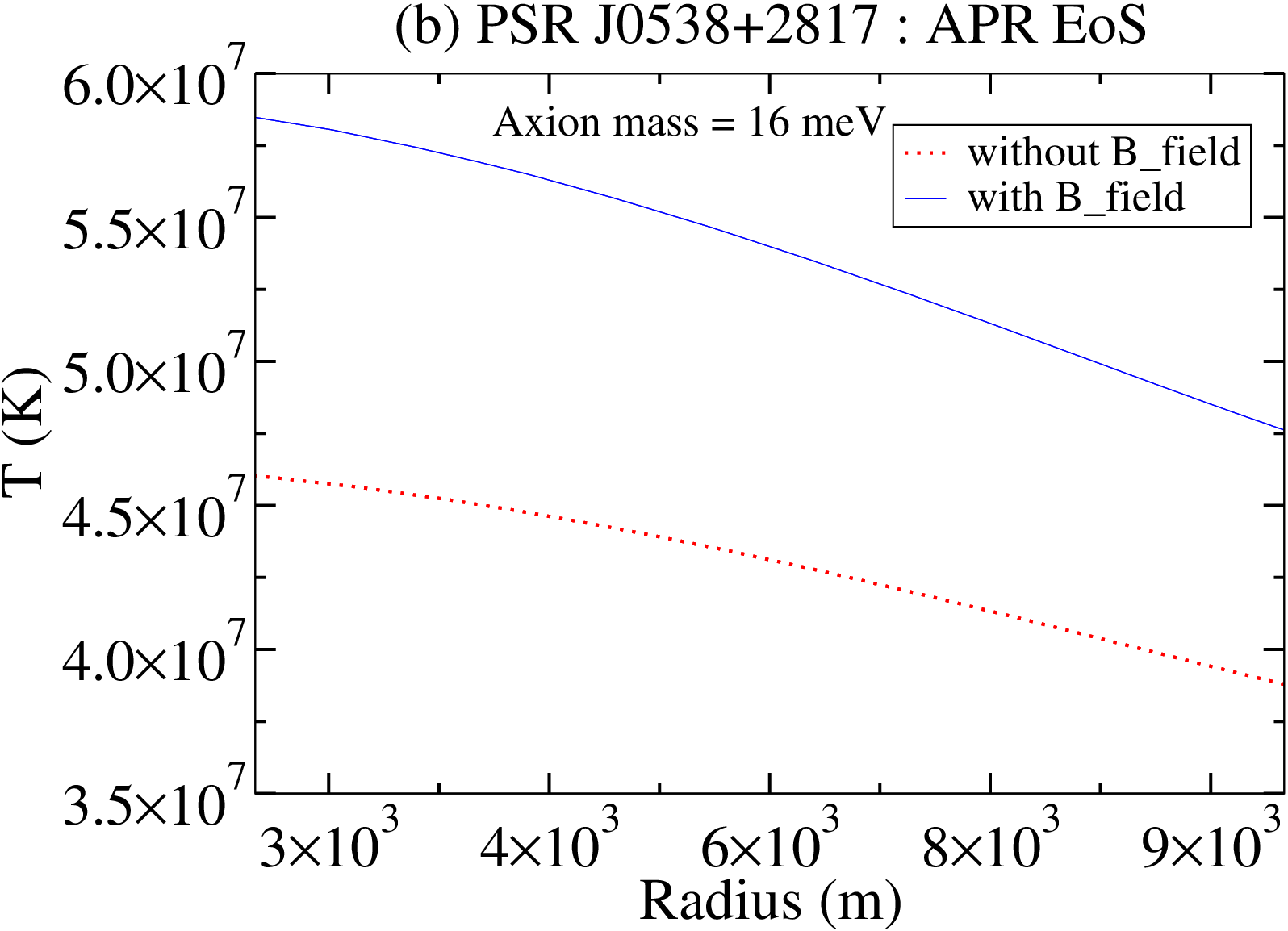} 
\end{tabular}
\end{figure*}

\begin{figure*}[htp!]
\begin{tabular}{cc}
\includegraphics[width=7.75cm]{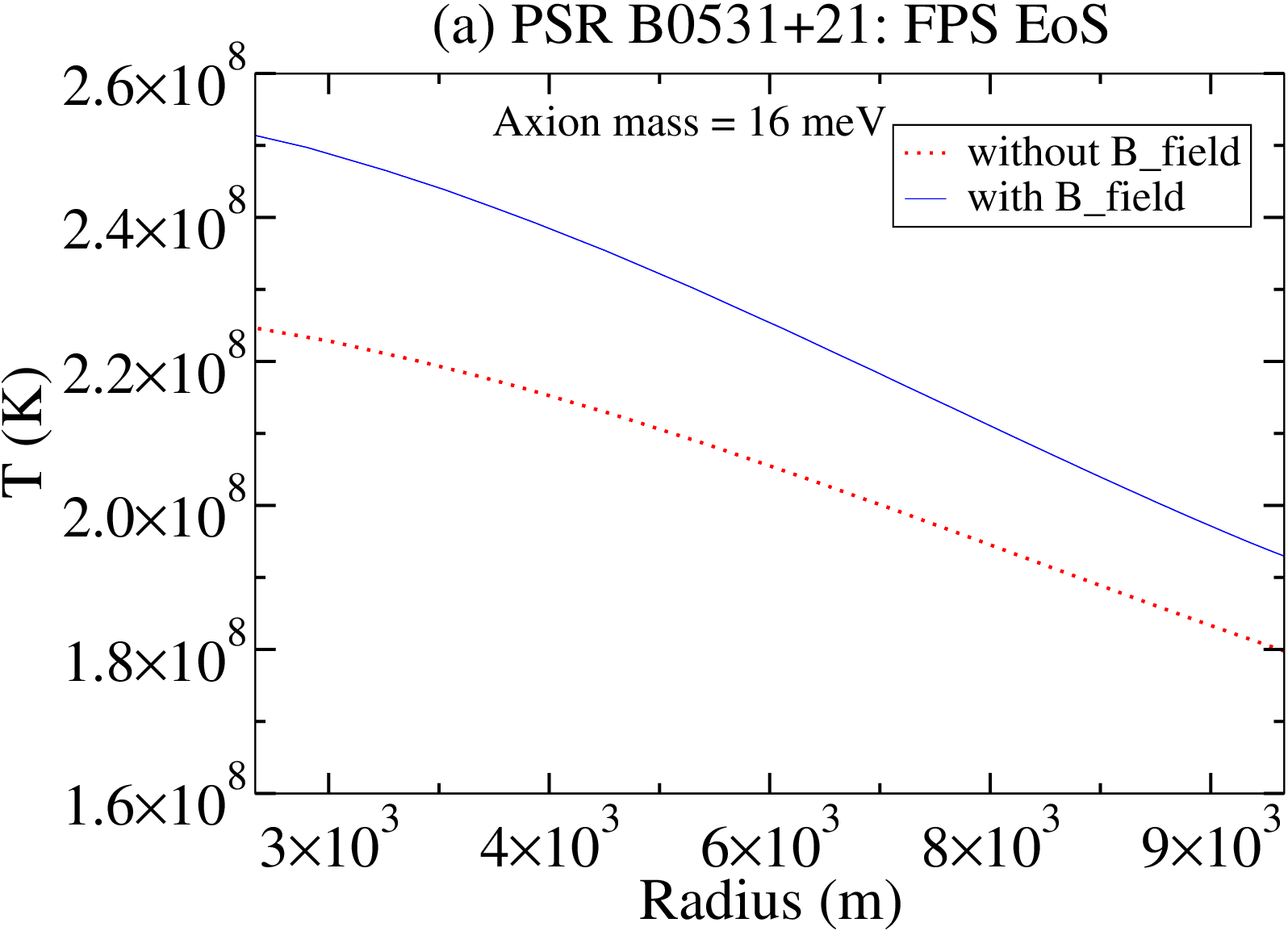}\ & \includegraphics[width=7.75cm]{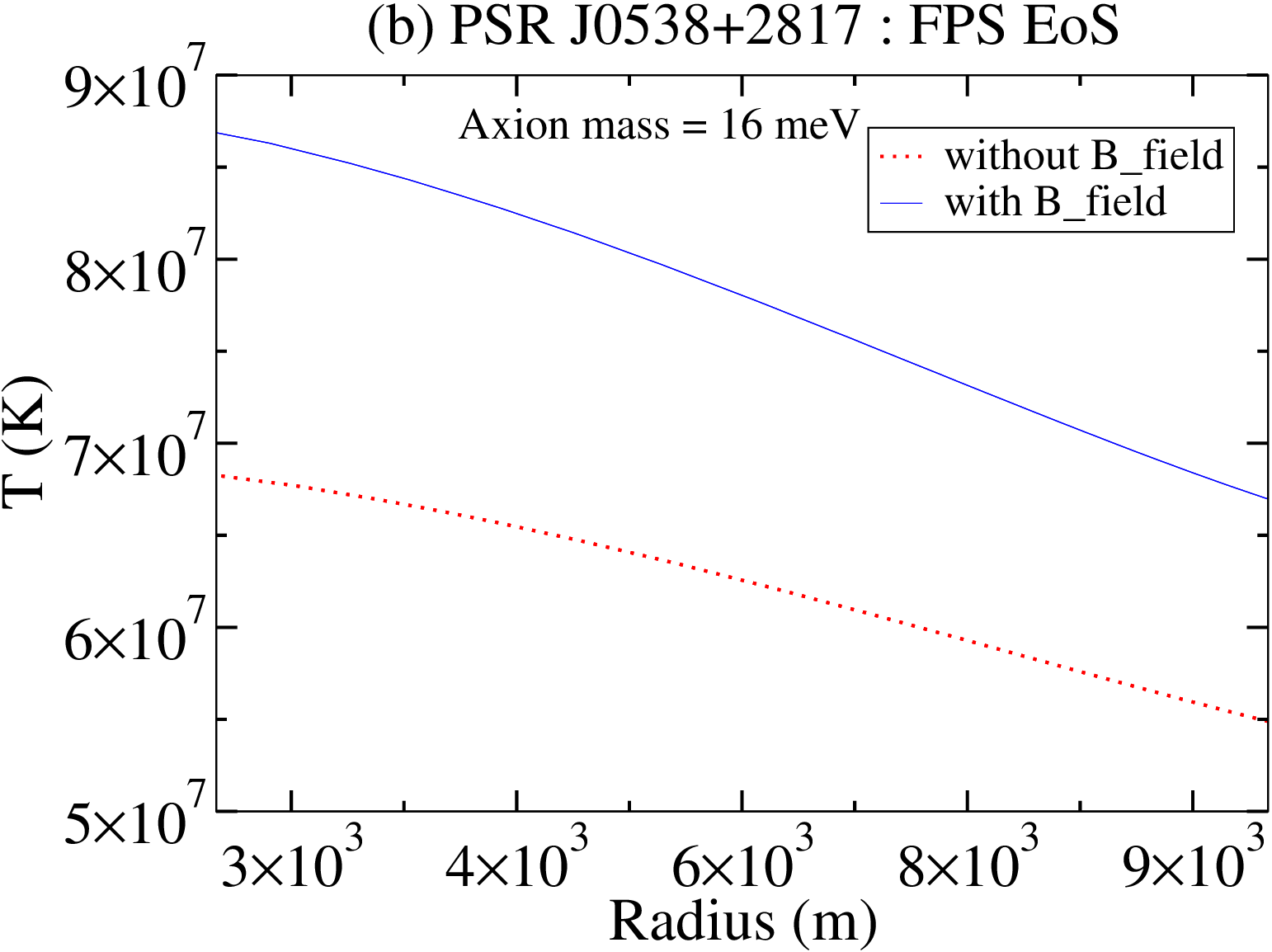} 
\end{tabular}
\end{figure*}
\begin{figure*}[htp!]
\begin{center}
\includegraphics[width=7.75cm]{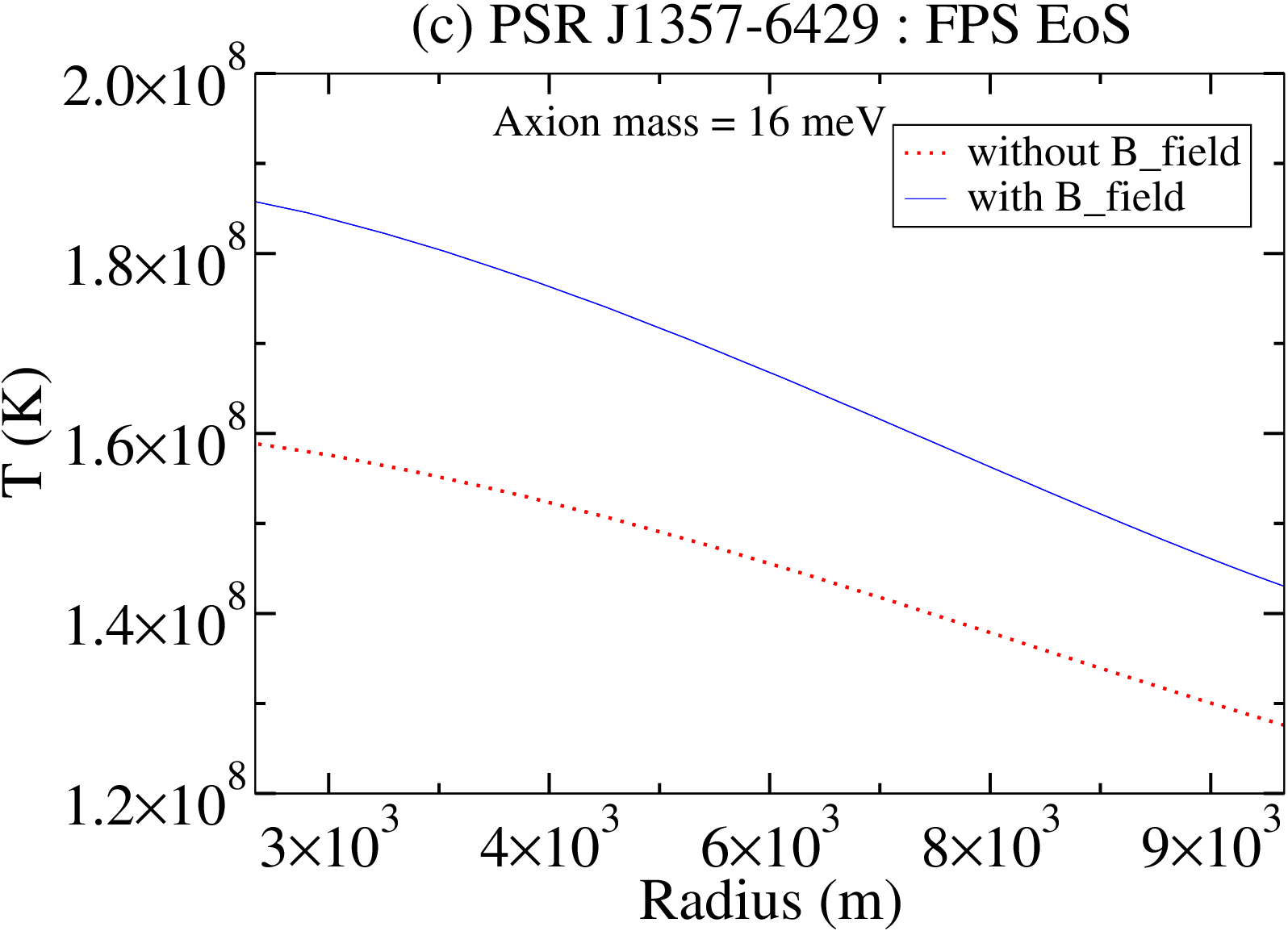}
\end{center} 
\caption{ The variation of temperature with radial distance $r$ for three different NSs for FPS EoS in the presence and absence of magnetic field.}   
\label{fig:nsradial2}
\end{figure*}

\begin{figure*}[htp!]
\begin{tabular}{cc}
\includegraphics[width=7.75cm]{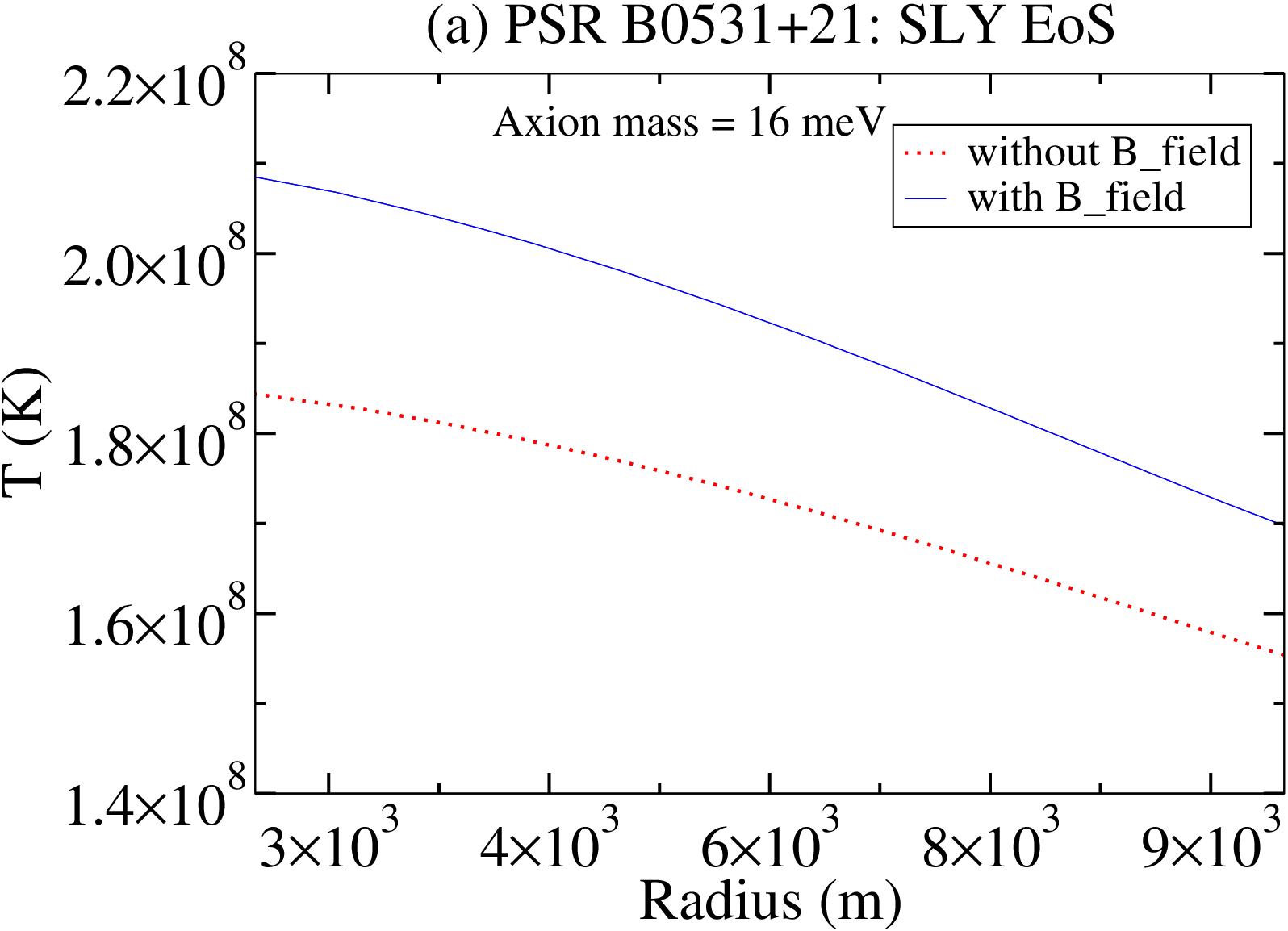}\ & \includegraphics[width=7.75cm]{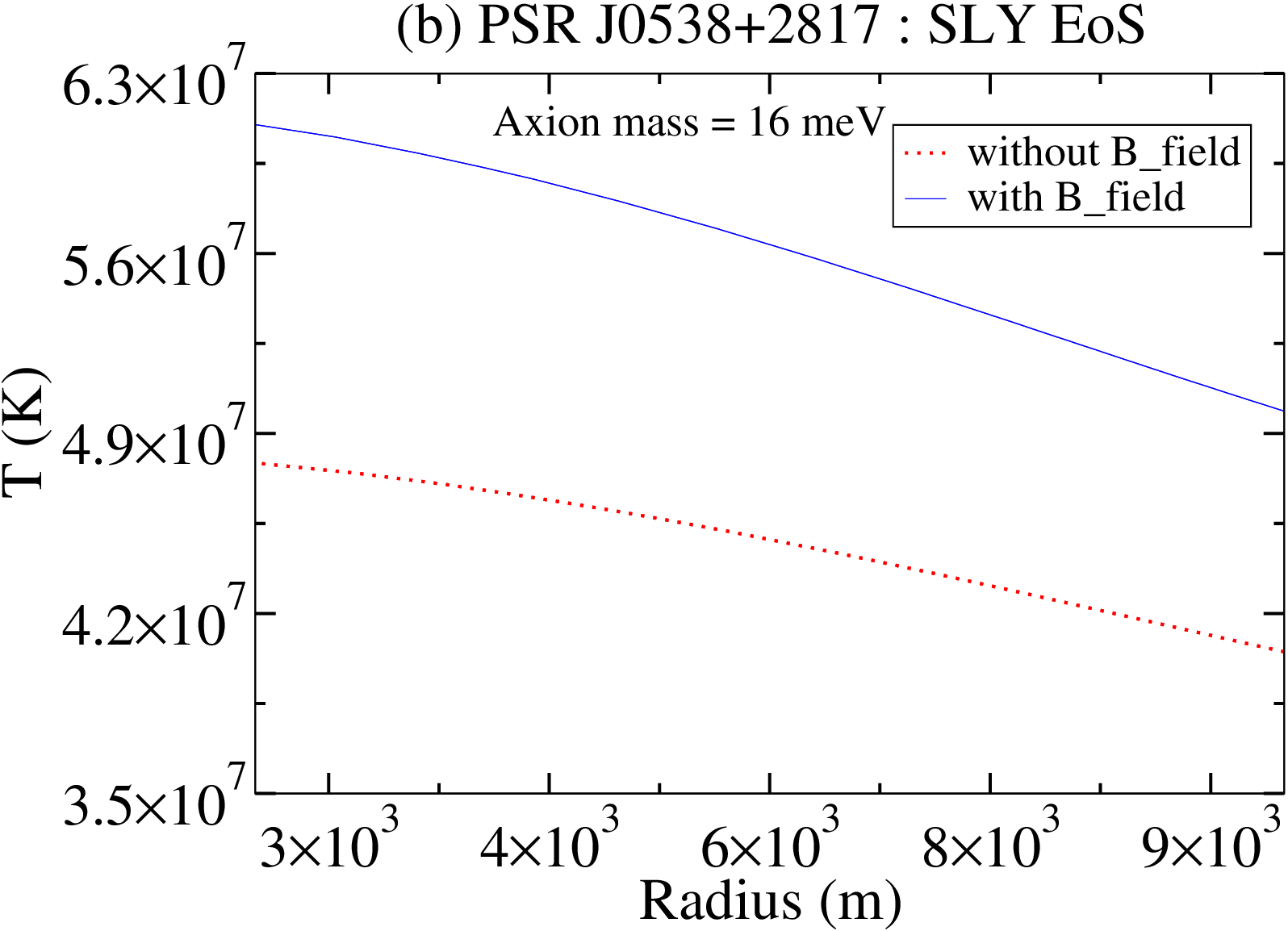} 
\end{tabular}
\end{figure*}
\begin{figure*}[htp!]
\begin{center}
\includegraphics[width=7.75cm]{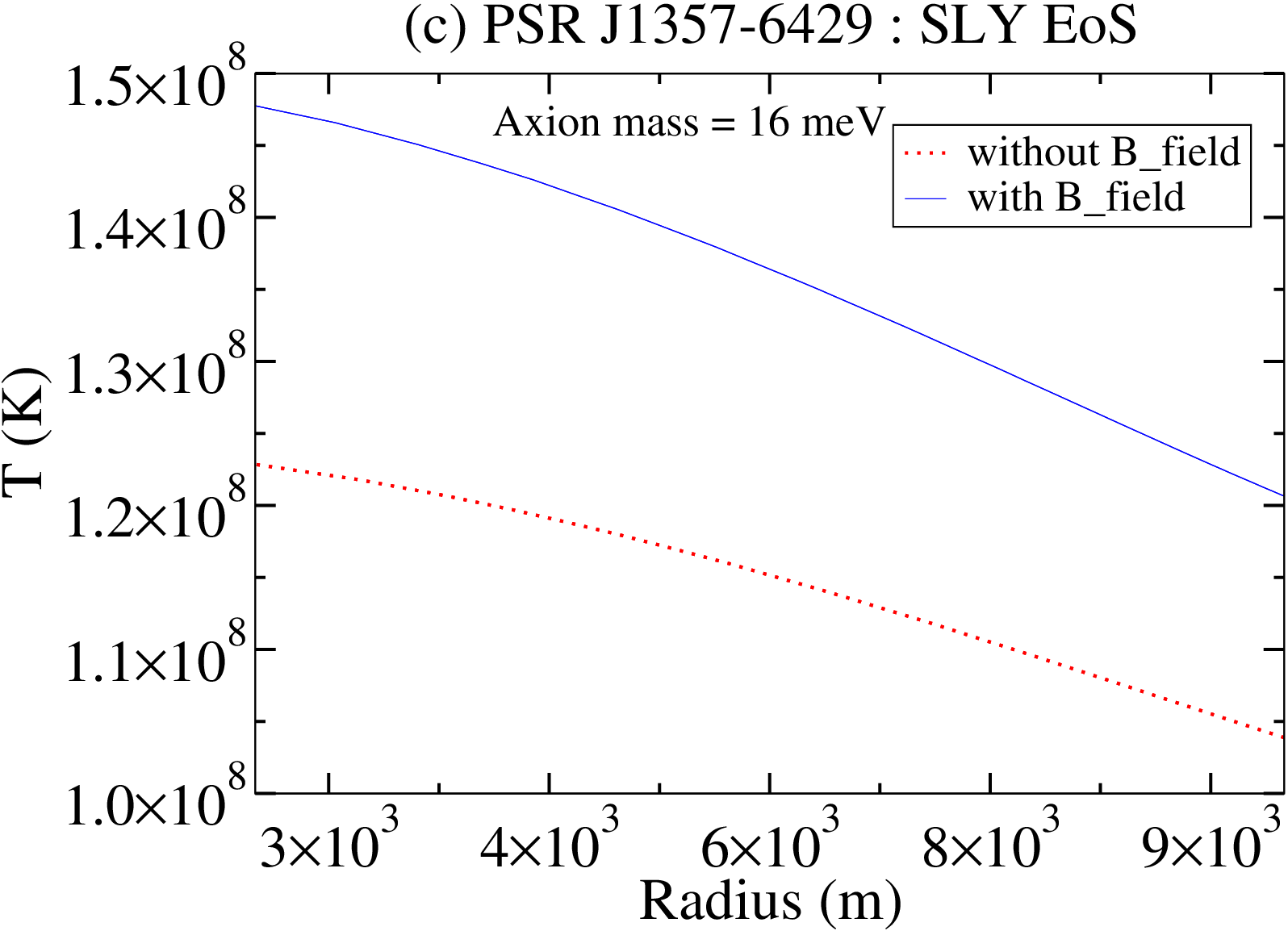}
\end{center}
\caption{ The variation of temperature with radial distance $r$  for three different NSs for SLY EoS in the presence and absence of magnetic field.}  
\label{fig:nsradial3}  
\end{figure*}
 Figure (\ref{fig:nsradial2}) and Figure (\ref{fig:nsradial3}) show the temperature as a function of the radius for the characteristic age of three NSs for FPS and SLY EoS, respectively. 
While the FPS EoS does not show any qualitatively different behavior from the APR EoS, the departure of the temperatures from the no-magnetic field case is always higher. The SLY EoS also indicates no qualitative changes in the temperature profile, and $T(r)$  at all radial distances remains higher than their counterpart in the no-magnetic field case at all times. At larger times  for this EoS, the temperature departure is higher than in the no-magnetic field. Table II shown below summarises the results.
\begin{widetext}
\begin{center}
\begin{table}[h]
\label{table:tablenos}
\begin{tabular}{ c | c |  c | c } 
 \hline
 \hline
 $m_{a}=0.016$ eV\ & Age $= 9.7\times 10^{2}$ yrs \ & Age $=3.3\times 10^{4} $ yrs \ & Age $=7.3 \times 10^{3} $ yrs \\ [0.5ex] 
 \hline
 APR \ & $10.5\%$\ & $20.8\%$\ & $16.1\%$ \\ 

 FPS \ & $9.8\%$\ & $20.6\%$\ & $13.7\%$\\

 SLY \ & $11.1\%$\ &  $21.1\%$\ & $16.5\%$\\ [1ex]
 \hline\hline 
\end{tabular}
\caption { Table showing the percentage departure of temperature from the magnetic field situation at $r=4500m$ for different EoS.}
\end{table}
\end{center}
\end{widetext}

\begin{widetext}
\begin{center}
\begin{table}[h!]
\label{tab:tabtwo}
\begin{tabular}{ c | c |  c  } 
 \hline \hline
 Age$= 9.7\times 10^{2}$ yrs, $T_{e}^{\infty}$ (observed)$= 1.55 \times 10^{6}$ (K) ~\cite{O'Dea_2014, refId0} \ & $T_{e}^{\infty}$ (without B  field) (K) \ & $T_{e}^{\infty}$ (with B field) (K) \\ [1ex] 
 \hline
 APR\ & $9.006 \times 10^{5}$\ & $1.020 \times 10^{6}$ \\ 

 FPS \ & $1.007 \times 10^{6}$\ & $1.130 \times 10^{6}$ \\
 
 SLY \ & $9.086 \times 10^{5}$\ & $1.018 \times 10^{6}$\\ 
 \hline\hline 
  Age $= 3.3\times 10^{4}$ yrs, $T_{e}^{\infty}$ (observed)$= 6.5 \times 10^{5}$(K) ~\cite{Xu_2021,Ng_2007} \ & $T_{e}^{\infty}$ (without B  field) (K) \ & $T_{e}^{\infty}$ (with B field) (K)\\ [1ex] 
 \hline
APR \ & $4.300 \times 10^{5}$\ & $5.052 \times 10^{5}$ \\ 

  FPS \ & $5.377 \times 10^{5}$\ & $6.277 \times 10^{5}$\\

  SLY \ & $4.419 \times 10^{5}$\ & $5.171 \times 10^{5}$\\ [1ex]
 \hline\hline 
 Age $= 7.3 \times 10^{3}$ yrs, $T_{e}^{\infty}$ (observed)$= 1  \times 10^{6} $(K) ~\cite{Zavlin_2007,2012ASPC..466...29K}  \ & $T_{e}^{\infty}$  (without B  field) (K) \ & $T_{e}^{\infty}$ (with B field) (K)\\ [1ex]  
\hline
  APR \ & $7.312 \times 10^{5}$\ & $8.409 \times 10^{5}$ \\ 

 FPS \ & $8.494 \times 10^{5}$\ & $9.573 \times 10^{5}$\\

 SLY\ & $7.385 \times 10^{5}$\ &$8.437 \times 10^{5}$\\ [1ex]
 \hline\hline 
\end{tabular}
\caption{ Table showing the comparison of observed surface temperature ($T_{e}^{\infty}$) for three NSs with  Age $= 9.7\times 10^{2}$ yrs, 
$3.3 \times 10^4$ yrs and $7.3 \times 10^3$ yrs, respectively with the theoretically predicted values for different EoS in presence/absence of magnetic field.}
\end{table}
\end{center}
\end{widetext}

\begin{figure*}[htp!]
\begin{center}
\includegraphics[width=7.75cm]{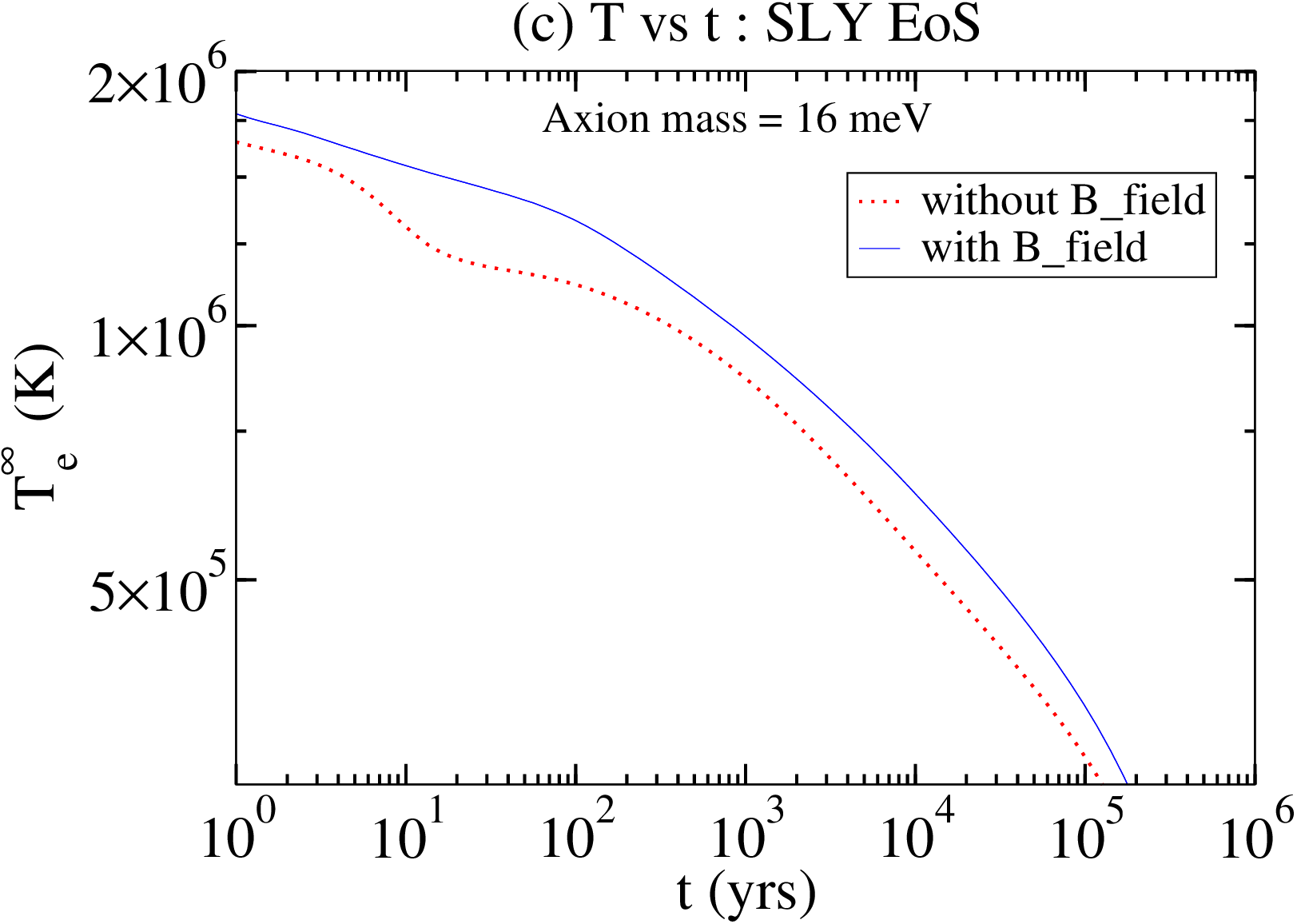}
\end{center} 
\caption{The variation of temperature with the time for three different EoS in the presence and absence of magnetic field.} 
\label{fig:timevar}
\end{figure*}

 In Figure (\ref{fig:timevar}a), we have shown the time evolution of the red shifted surface temperature of the NS by incorporating the effect of magnetic field for axion mass $\sim 16$ meV, assuming that the NS matter satisfies the APR EoS. The variation of temperature in the absence of a magnetic field is also shown for comparison. In the presence of a magnetic field, the temperature versus time curve lies above the corresponding variation curve without the magnetic field for the APR EoS. Between time  $10^1$ yrs and $10^2$ yrs, the separation increases and subsequently, both curves have shown the same qualitative behavior throughout the entire cooling time duration. However, the temperature in the presence of magnetic field always remain higher than the temperatures without magnetic field at all times. We have shown the comparative behavior of the cooling curves for two other EoSs as well, namely, FPS and SLY,  for the same axion mass  $\sim 16$ meV as shown in Figure (\ref{fig:timevar}b) and Figure (\ref{fig:timevar}c), respectively.\\

Figure (\ref{fig:timevar}b), shows a significant difference in the cooling curves up to  $10^2$ yrs. The separation decreases beyond  $10^5$ yrs and the difference in the temperature for the case with magnetic field and without magnetic field becomes maximal at $ \sim 4 \times 10^1$ yrs.
The comparison of the predicted and the observed red shifted surface temperatures for the three NSs are summarized in Table (III). It is worthwhile to mention here that our predicted values of surface temperatures  with magnetic field are closer to the corresponding observed values for all three NSs. However, strong sensitivity with respect to the EoS implies that the appropriate fine-tuning is needed in the presence of the magnetic field.

\begin{figure*}[htp!]
\begin{tabular}{cc}
\includegraphics[width=7.75cm]{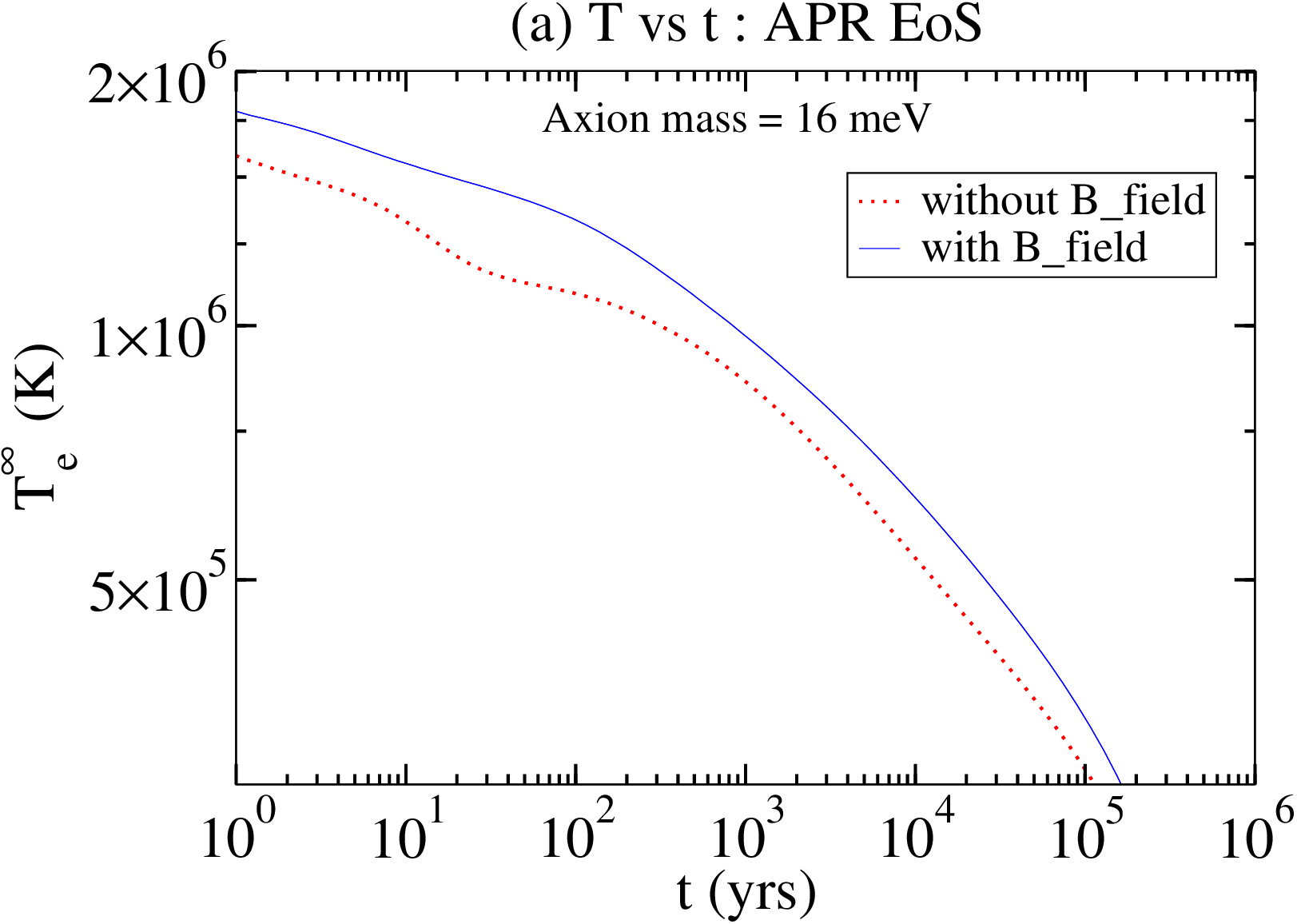}\ & \includegraphics[width=7.75cm]{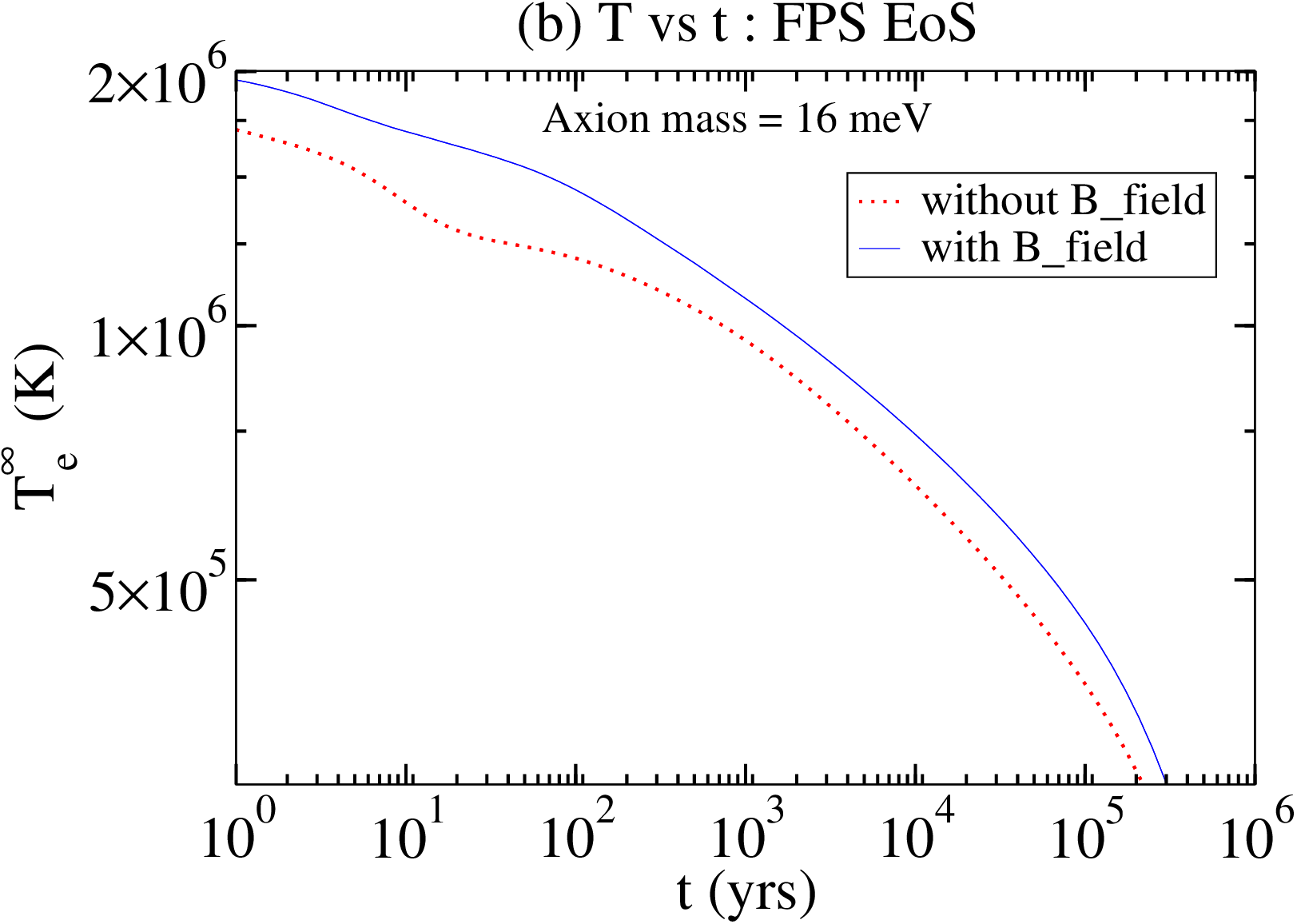} 
\end{tabular}
\end{figure*}
At the early times, the cooling curve shows a significant difference between the results in the presence and absence of magnetic fields. From Figure (\ref{fig:timevar}c), it is seen that there is a noteworthy difference between the two curves up to time  $10^2$ yrs. Subsequently, the separation decreases, and beyond time $10^2$ yrs, the difference between the two curves remains almost constant and decreases only at later times. 

 The EoS of the matter present inside the NS also affects the cooling profile's modification in the presence of magnetic fields. While the overall qualitative features remain the same. There are significant EoS-sensitive departures which indicates that the modeling of NSs with strong magnetic fields would have an imprint of the magnetic field distribution. This behavior may affect the observable properties of NSs significantly. Finally, we have studied the luminosities of neutrinos, axions, and photons as a function of time in the absence and presence of the magnetic field for three EoSs. We expect to see the imprint of the strong magnetic field in these results.\\

\begin{figure*}[htb!]
\begin{tabular}{cc}
\includegraphics[width=7.75cm]{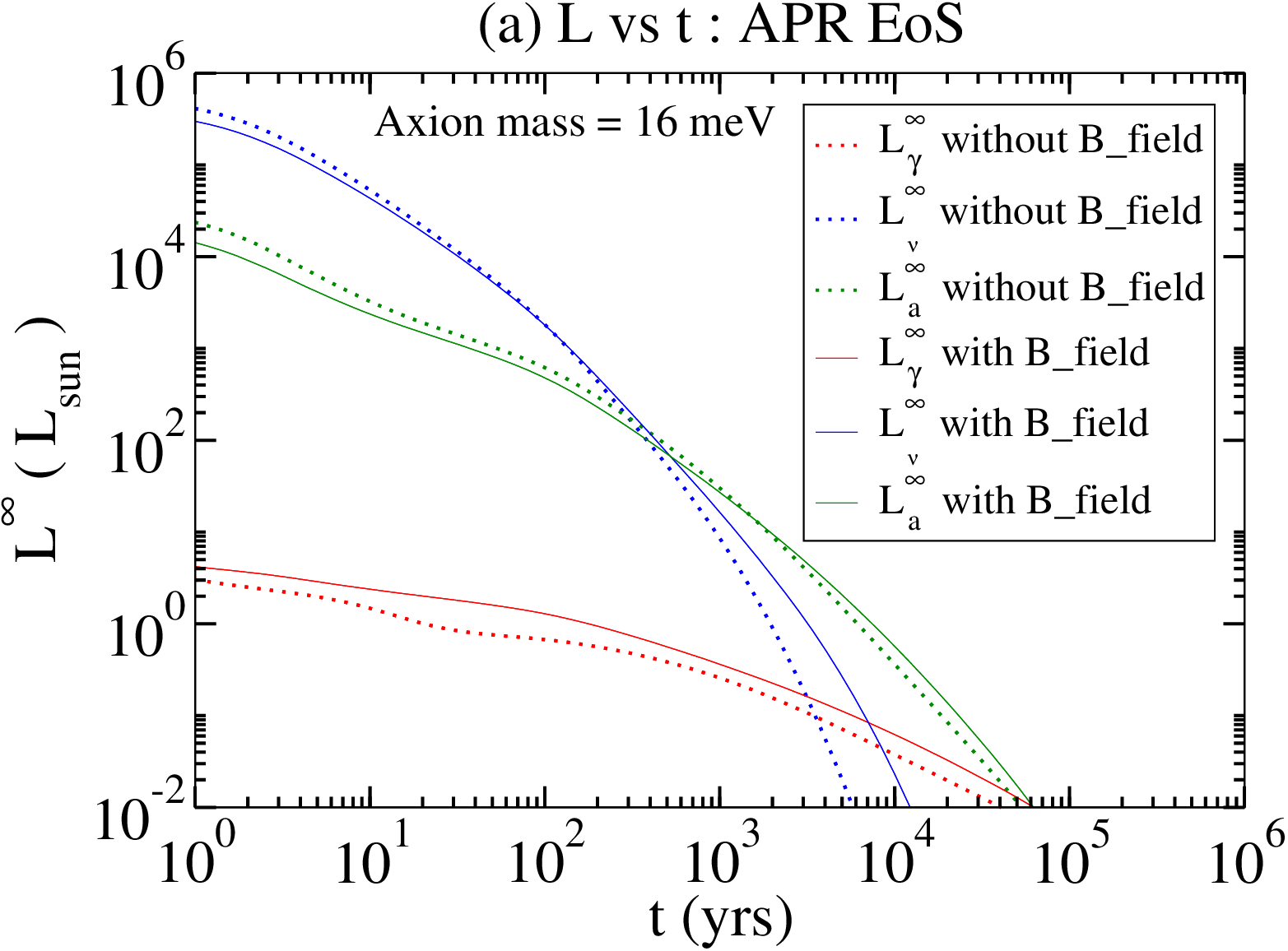}\ & \includegraphics[width=7.75cm]{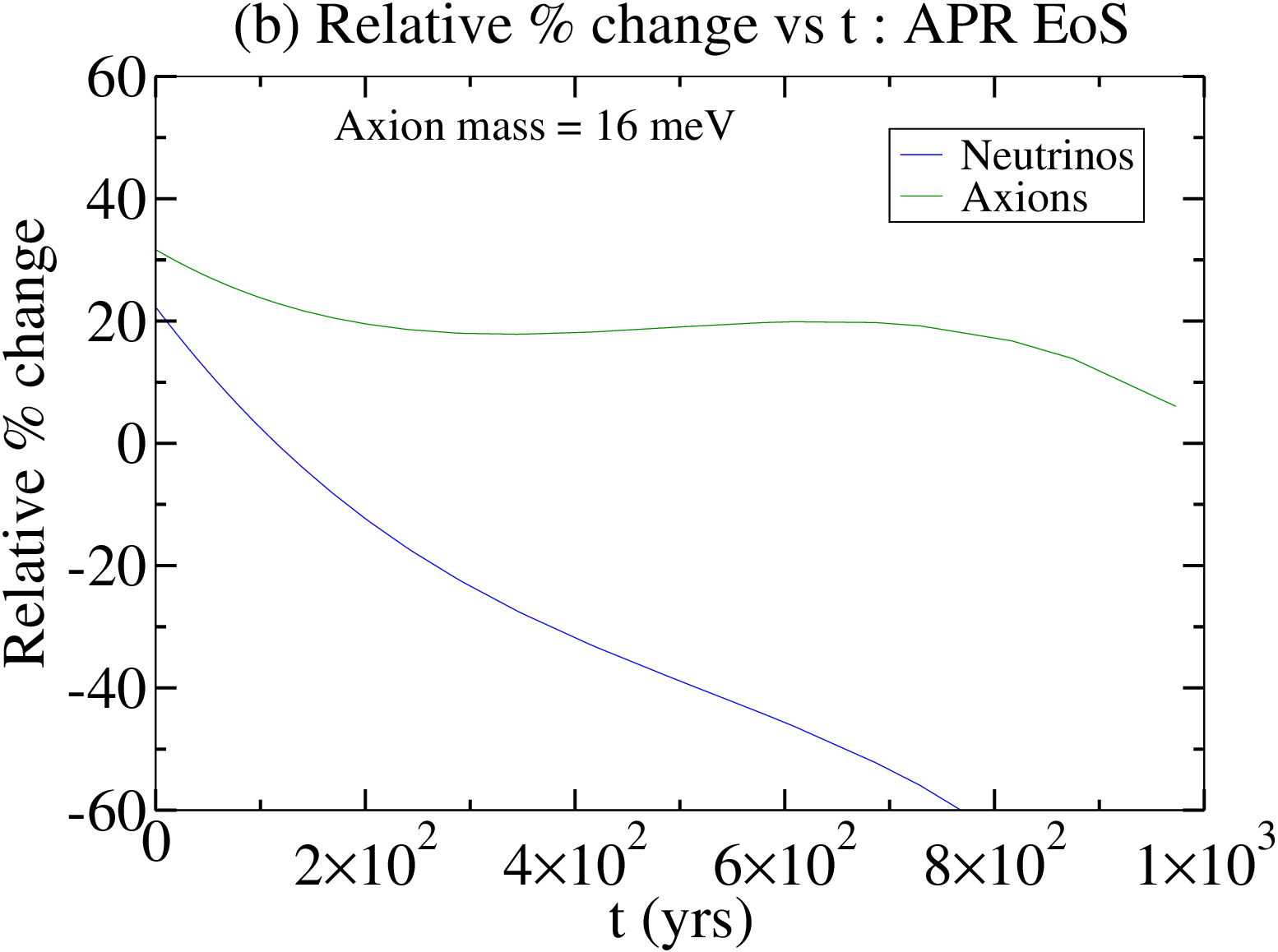}
\end{tabular}
\caption{ The variation of luminosity with the time of axion mass  $16$ meV, for APR EoS (a) in the presence and absence of magnetic field. The relative $\%$ changes in axions \& neutrino luminosity with time of axion mass  $16$ meV, for APR EoS (b) (Best fit curve).}  
\label{fig:ltimevar1}
\end{figure*}

 Figure (\ref{fig:ltimevar1}a) depicts the luminosity versus time at an axion mass $16$ meV for APR EoS.
The luminositites are all expressed in units of solar luminosity $L_{\odot} = 3.826 \times 10^{33}\, {\rm ergs /s}$.
The photon luminosity~\cite {2015SSRv..191..239P} at very short times are not accurate as our time unit is coarse and extrapolated.
The incomplete modeling of NS due to uncertainty in EoS may also be responsible for the lower estimate of luminosities at short times.
The neutrino luminosity dominates over the photon and axion luminosity at early times. The presence of magnetic fields does not change the qualitative behavior, although there is a significant difference in the neutrino luminosity. The magnitude of the luminosity of photons,  axions and neutrinos is quite sensitive to the axion mass and also shows a significant departure in the presence of the magnetic field.  The time evolution of photon, neutrino and axion luminosities in the absence of magnetic field closely resembles the results in an earlier work~\cite{Buschmann:2021juv}.

 The expression to calculate the relative percentage change is given by 
relative $\%$ change $= (\frac{L_{WB}-L_{B}}{L_{WB}})\times 100
$, where,  $L_{WB}$ and $L_{B}$ correspond to the luminosities in the absence and presence of a magnetic field, respectively. Figure (\ref{fig:ltimevar1}b) shows  the relative percentage changes in axions and neutrino luminosity as a function of time due to the impact of the magnetic field for the APR EoS. The relative percentage change in axions luminosity dominates over the neutrino luminosity at all times.  We find that the relative percentage change in axion luminosity remains roughly constant in time as opposed to the relative percentage change in neutrino luminosity.

\begin{figure*}[htb!]
\begin{tabular}{cc}
\includegraphics[width=7.75cm]{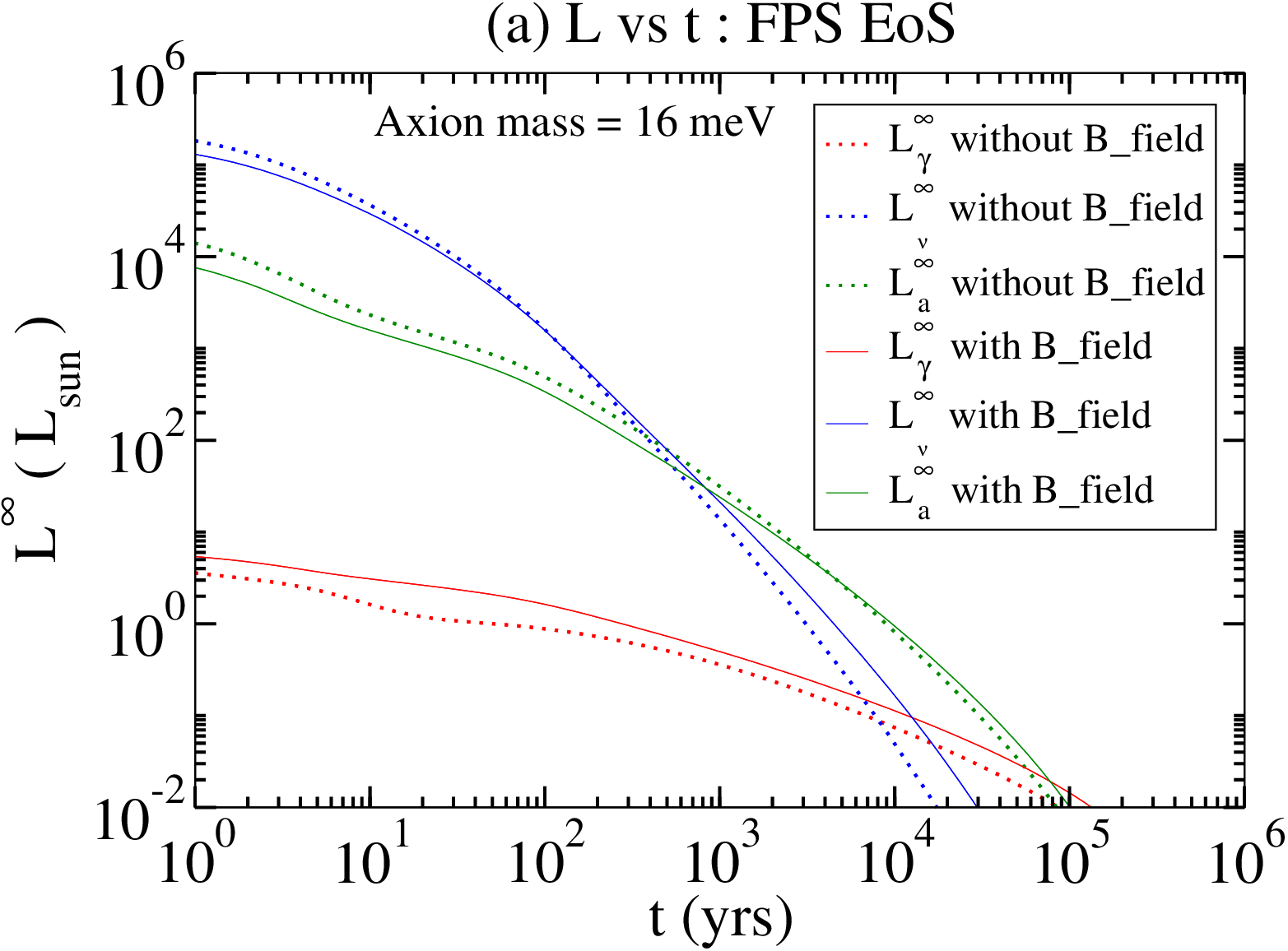}\ & \includegraphics[width=7.75cm]{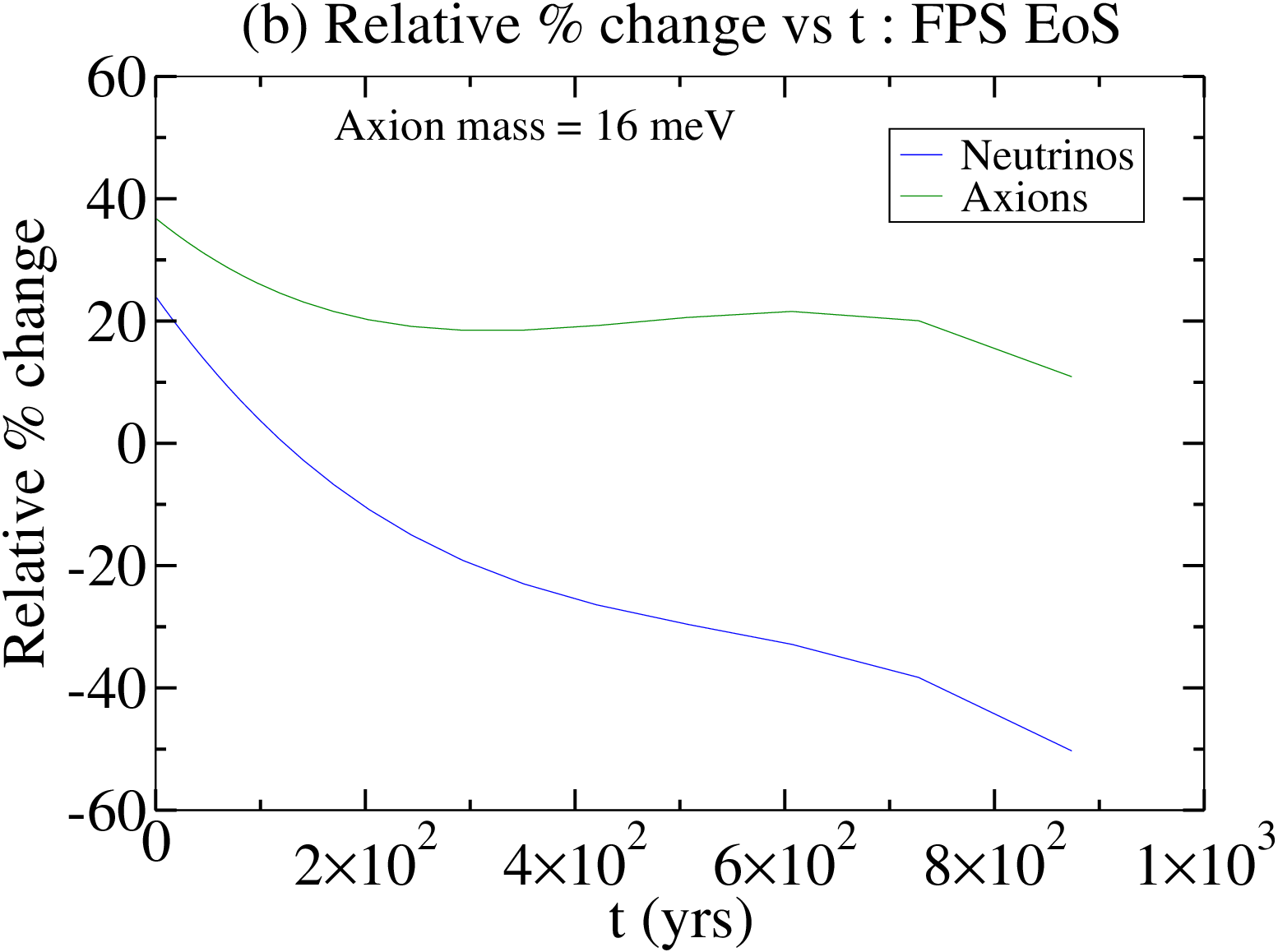}
\end{tabular}
\caption{The variation of luminosity with the time for axion mass  $16$ meV, for FPS EoS (a) in the presence and absence of magnetic field. The relative $\%$ changes in axions $\&$ neutrino luminosity with time for axion mass $16$ meV, for FPS EoS (b) (Best fit curve).}  
\label{fig:ltimevar2}
\end{figure*}

Figure (\ref{fig:ltimevar2}a) shows the luminosity versus time at an axion mass $16$ meV for the FPS EoS. The qualitative behavior does not seem to change much with the magnetic field, and a significant departure is seen in the neutrino luminosity and axion luminosity, respectively, in the presence of the magnetic field. Figure (\ref{fig:ltimevar2}b) depicts the relative percentage changes in axions and neutrino luminosity versus time due to the magnetic field at an axion mass $16$ meV for FPS EoS. Here also we  find that the relative percentage change in axion luminosity remains more or less constant in time as opposed to the relative $\%$ change in neutrino luminosity.\\

\begin{figure*}[htb!]
\begin{tabular}{cc}
\includegraphics[width=7.75cm]{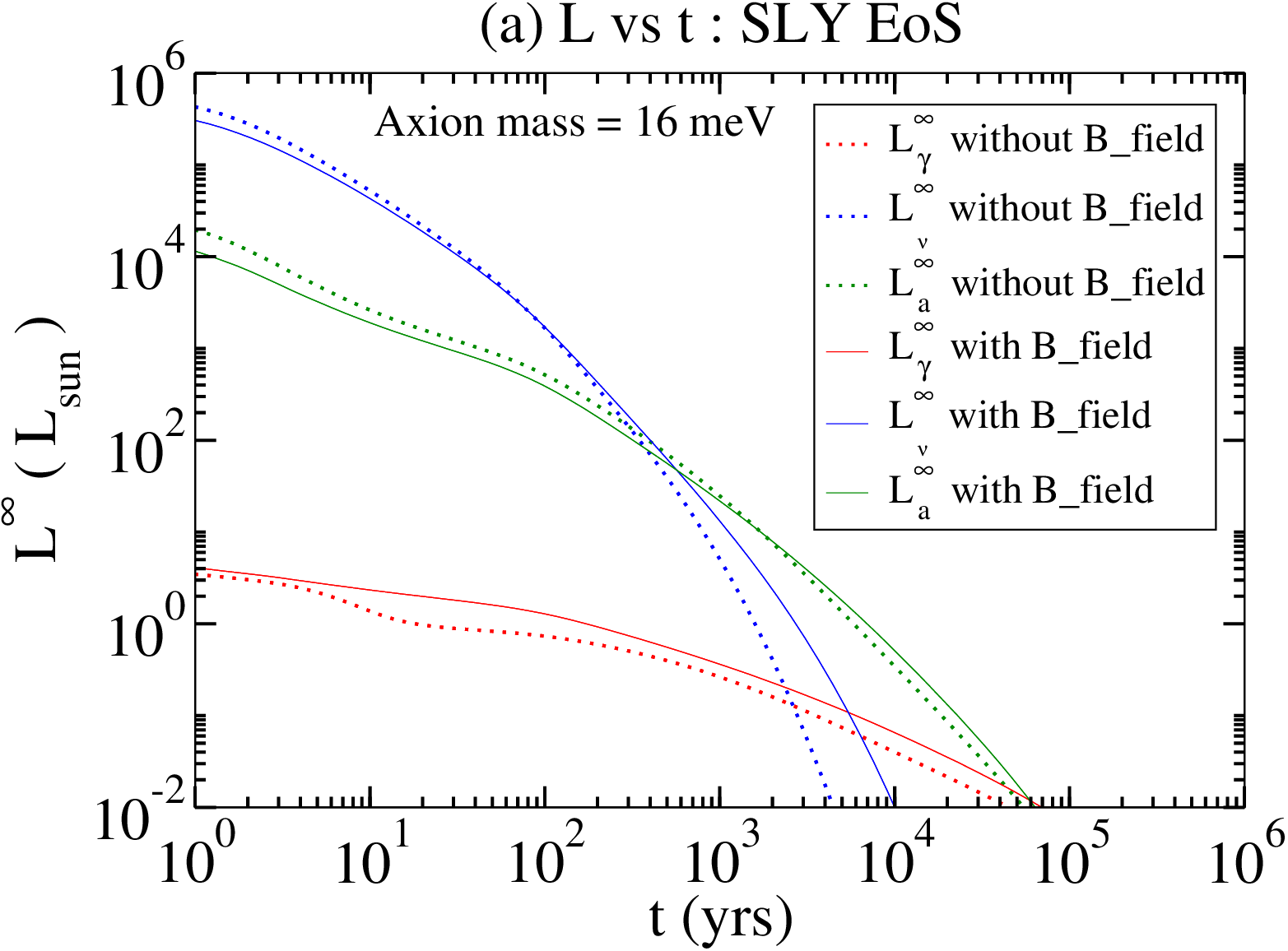}\ & \includegraphics[width=7.75cm]{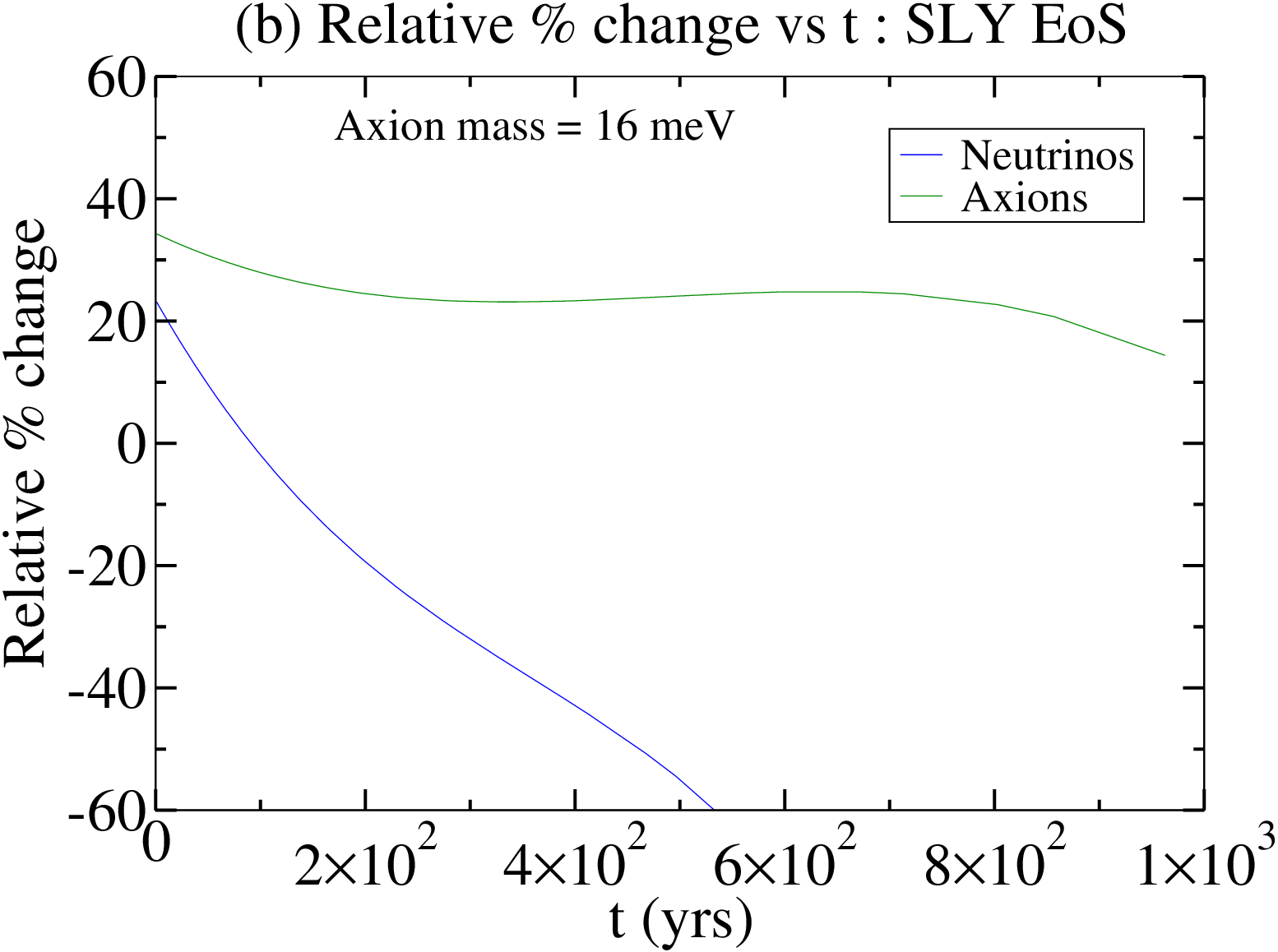}
\end{tabular}
\caption{The variation of luminosity with the time for axion mass $16$ meV, for SLY EoS (a) in the presence and absence of magnetic field. The relative $\%$ changes in axions $\&$ neutrino luminosity with the time for axion mass  $16$ meV, for SLY EoS (b) (Best fit curve).}   
\label{fig:ltimevar3}
\end{figure*}

 In Figure (\ref{fig:ltimevar3}a), we have presented the luminosity versus time at an axion mass $16$ meV for SLY EoS. Here, also a large departure is seen in the neutrino and axion luminosity, respectively, in the presence of the magnetic field. Figure (\ref{fig:ltimevar3}b) depicts the variation of relative percentage changes in axions and neutrino luminosity with time due to magnetic field effect at an axion mass $16$ meV, for SLY EoS. 
 We see a qualitative behavior similar to the APR case. \\
In our current work, the axion mass bound is obtained for APR EOS by considering the surface temperature of three NSs (\cite{O'Dea_2014, refId0,Xu_2021,Ng_2007,Zavlin_2007,2012ASPC..466...29K}).
 We find the average best fit bound on the axion mass from these measurements  
 in the model without a magnetic field is around $\sim 5-6$ meV. 
 Including the magnetic field effect, we obtain the bounds as $\sim 9$ meV, which is much closer to the recently reported axion mass bound of $10$ meV and $16$ meV than when no magnetic field is applied. Axion mass bounds for the other two EOSs, namely FPS and SLY, almost lie within the above-mentioned limits.
 The role of the magnetic field is thus crucial in explaining the observed neutron star thermal properties.
 The mass bounds are expected to have degeneracies with the EoS. While this requires closer study, a cursory analysis indicates that the EoS affects the mass and pressure profiles, which implicitly affects the thermal history of the star. Further, the EoS also affects the value of $\bar r$ and changes the magnetic field profile slightly and thus has an effect on the axion mass bounds. \\

\section{Conclusion and Future Outlook}
\label{conc}
In the present work, we investigate the variations in luminosities of observables (neutrinos, axions, and photons) in the presence of strong magnetic fields for highly magnetized NSs. We have also explored the associated cooling rate characterized by the temperature versus time and radius, respectively. Our investigations are based on the assumption that axions, neutrinos, and photons constitute the emissions from the core and crust of NSs~\cite{Buschmann:2019pfp}.  We have also shown the relative percentage change in axion and neutrino luminosity due to the effect of the magnetic field. Further, we have summarized our findings below:

\begin{itemize}
 \item The emission characteristics of NSs are subject to intricate influences stemming from factors such as magnetic fields, radiation from nearby objects, and significant gravitational effects from astrophysical entities. Remarkably, the magnetic fields within NSs considerably impact internal structures, affecting parameters like mass and pressure profiles. These magnetic fields may extend their influence to the equation of state (EoS), although their effectiveness under the extreme baryon density conditions within NSs remains uncertain. Nevertheless, the magnetic field distinctly alters thermal and other macroscopic properties.

 \item While current study did not incorporate the magnetic field's influence on EoSs of core matter. We employed a modified set of Tolman-Oppenheimer-Volkoff (TOV) equations~\cite{2019PhRvC..99e5811C} along with three distinct EoSs namely: APR, FPS, and SLY to obtain mass and pressure profiles for NSs. The obtained profiles are used to compute cooling rates and luminosities using the NSCool code. This code, incorporating axion emission from the NS core and crust, assumed spherically symmetric NSs with a perfectly isothermal core and excluded heating mechanisms.

 \item Our analysis encompassed the calculation and plotting of luminosities for neutrinos, axions, and photons, both with and without magnetic fields, considering axion mass of $16$ meV for each EoS. Notably, the cooling and luminosity data have been plotted with a fixed mean radius, accounting for deformation due to the magnetic field. Due to the magnetic field, the relative percentage change in axion luminosity remains higher than the neutrino luminosity at all times for all three EOSs.

 \item Our findings unequivocally reveal a substantial influence of magnetic fields on luminosities and cooling rates. However, it is imperative to acknowledge that this influence is EoS-dependent, and its uncertainties are particularly pronounced under extremely high-density and low-temperature conditions. The investigation into axion-to-photon conversion in these highly magnetized NSs remains a subject for future research, as we deferred its exploration in this current study. It is worth mentioning here that due to the magnetic field, the axion mass bound increases slightly compared to without a magnetic field. 
\end{itemize}
 
 \vskip 1cm
\section*{Acknowledgments}
We thank the anonymous referee for carefully
reading the manuscript and for the remarks that
helped us a lot in improving the quality of our work. We are grateful to Dany Page, Malte Buschmann, and T. Opferkuch for clearing our doubts about the NSCool code. The author (SY) acknowledges Birla Institute of Technology and Science, Pilani, Pilani Campus, Rajasthan, for financial support.  

\bibliography{MS_CRSV2}  
\newpage
\section*{Appendix}
\subsection{Variation of luminosity  for different axion masses for three different EoS}       
\begin{figure*}[htb!]
\begin{tabular}{cc}
\includegraphics[width=7.5cm]{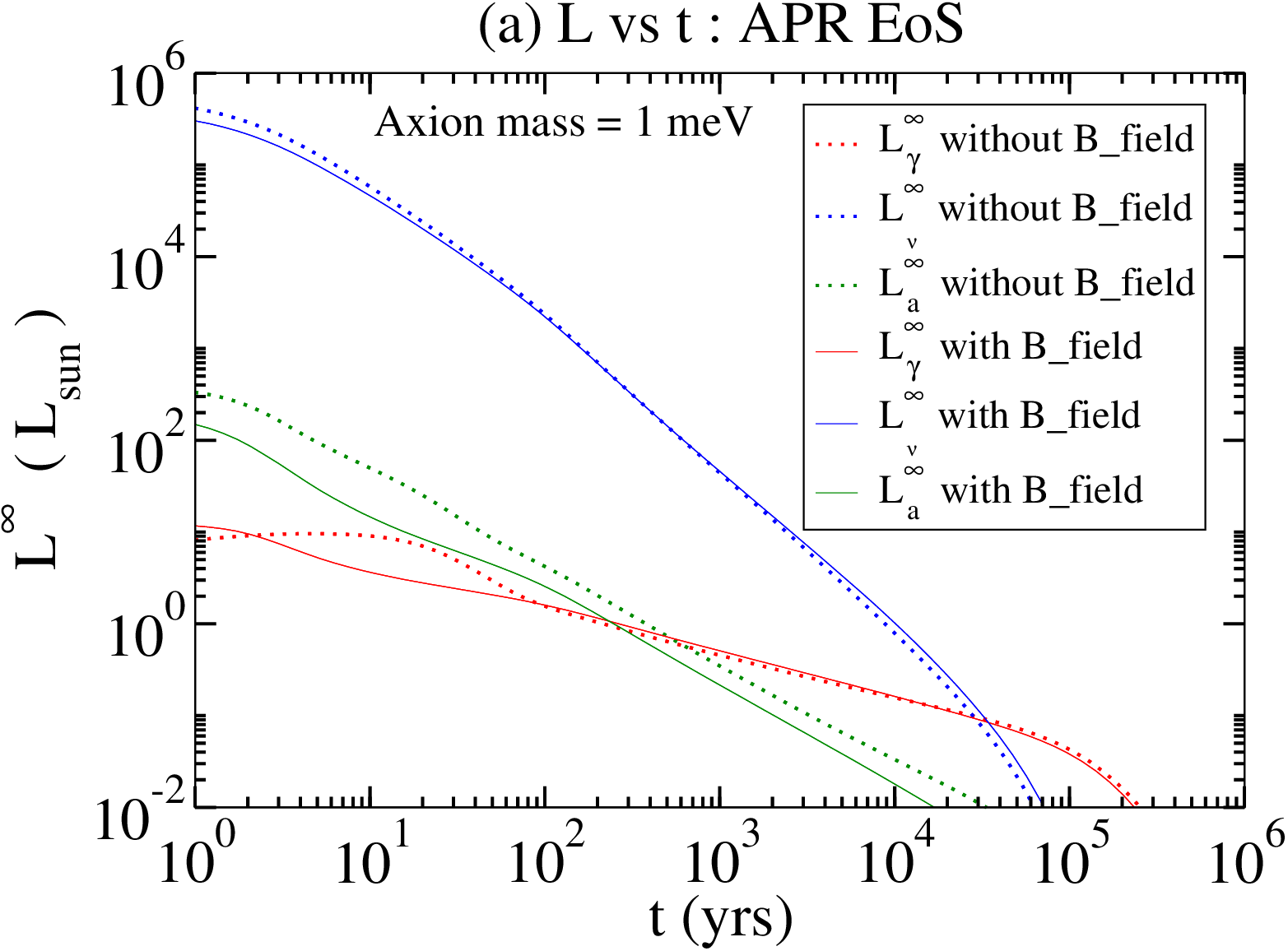} & \includegraphics[width=7.5cm]{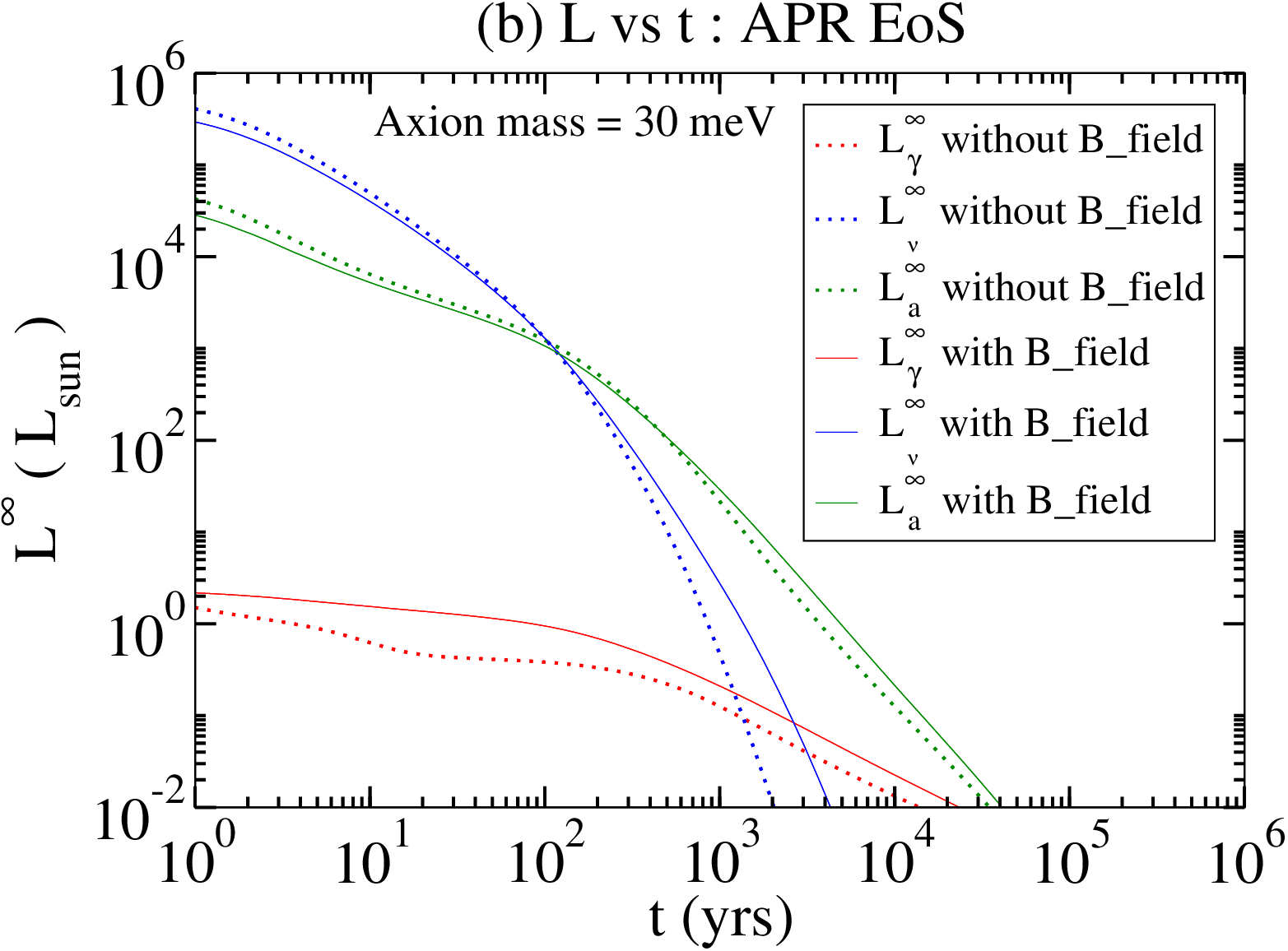}
\end{tabular}
\caption{ The variation of luminosity with the time of two different axion masses $1$ meV \& $30$ meV, respectively, for APR EoS in the presence and absence of magnetic field.}   
\label{fig:ltimevar5}
\end{figure*}

\begin{figure*}[htb!]
\begin{tabular}{cc}
\includegraphics[width=7.5cm]{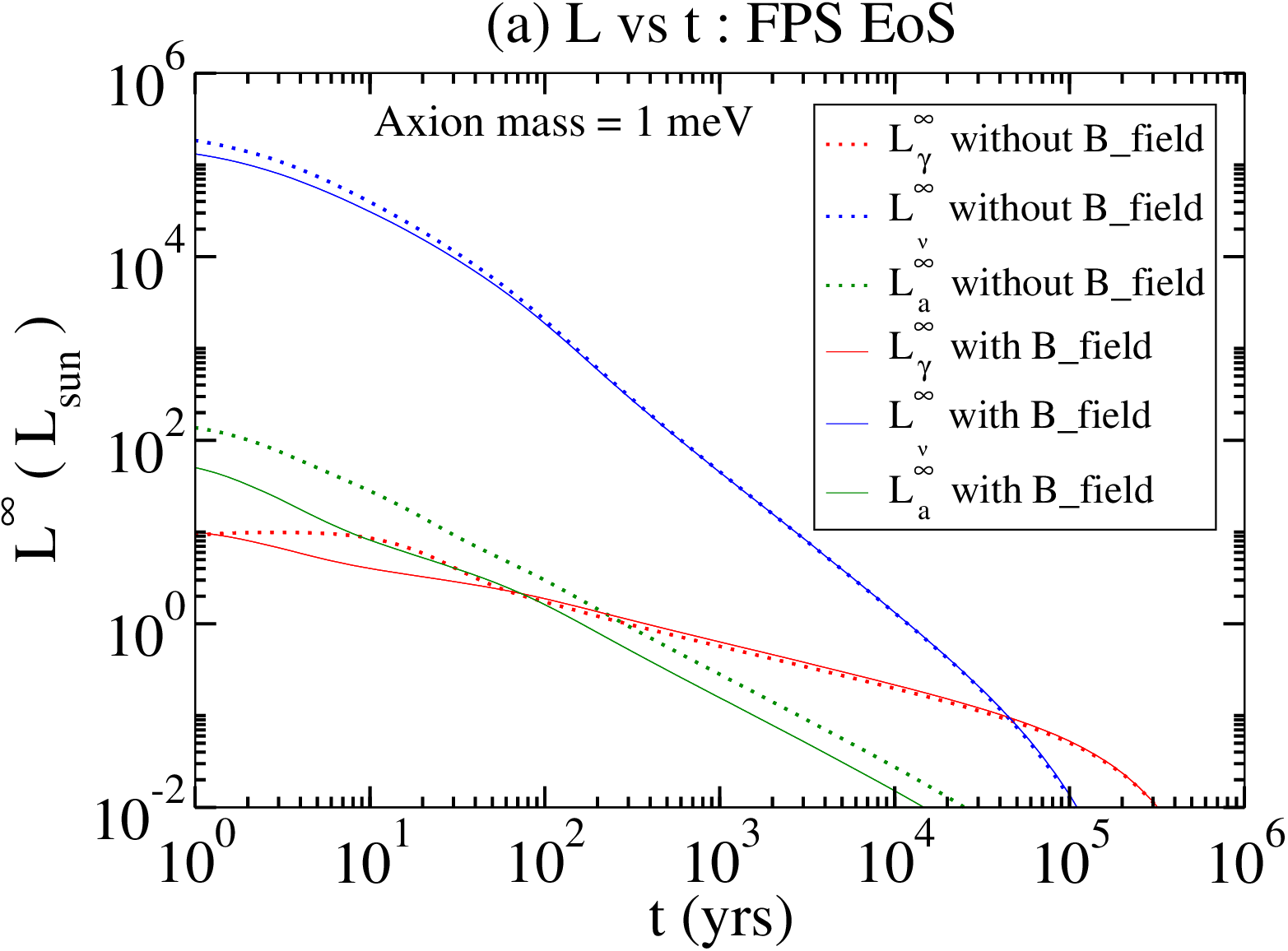} & \includegraphics[width=7.5cm]{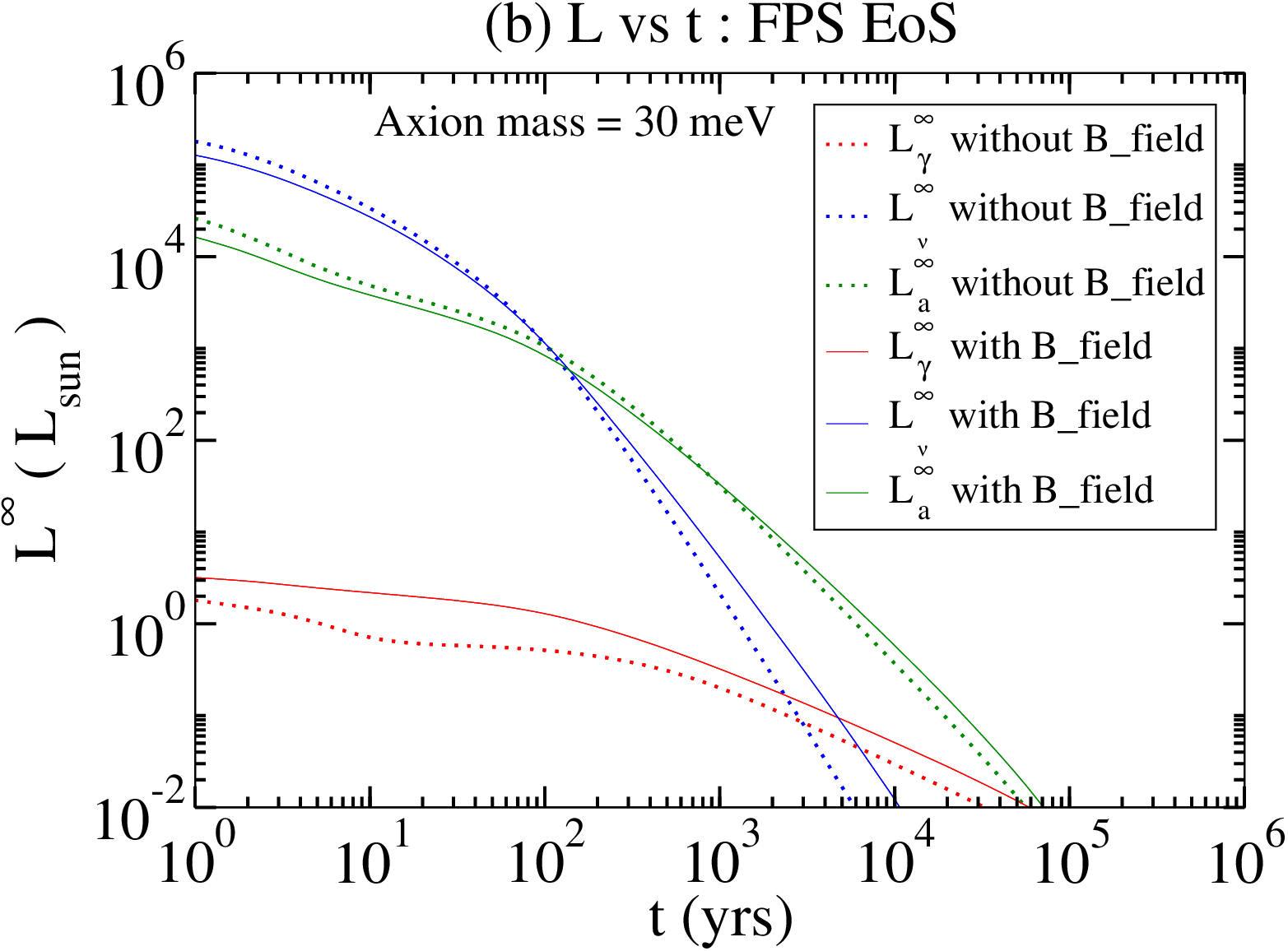}
\end{tabular}
\caption{The variation of luminosity with the time of two different axion masses $1$ meV, \& $30$ meV respectively for FPS EoS in the presence and absence of magnetic field.}  
\label{fig:ltimevar6}
\end{figure*}

\begin{figure*}[htb!]
\begin{tabular}{cc}
\includegraphics[width=7.5cm]{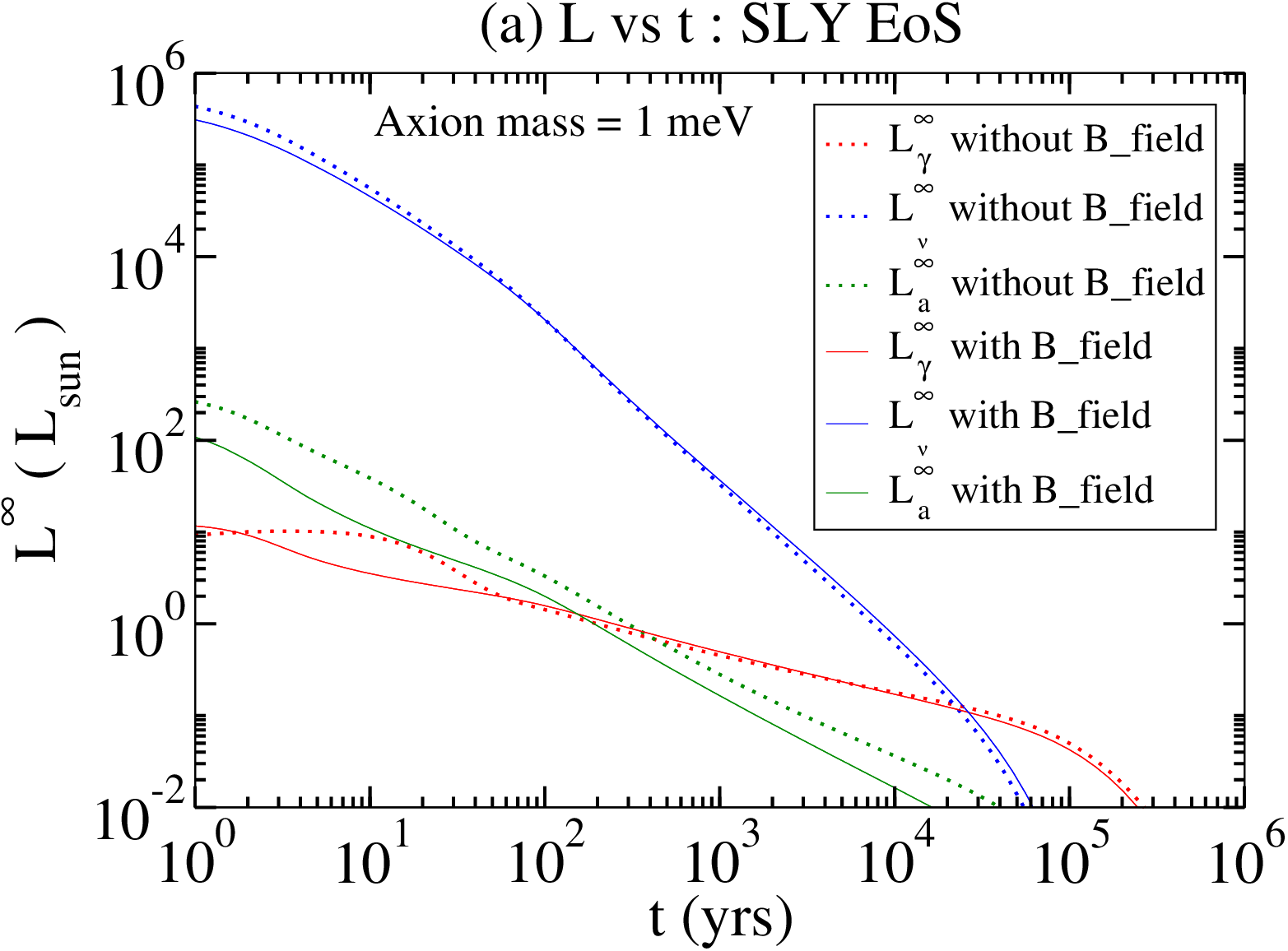} & \includegraphics[width=7.5cm]{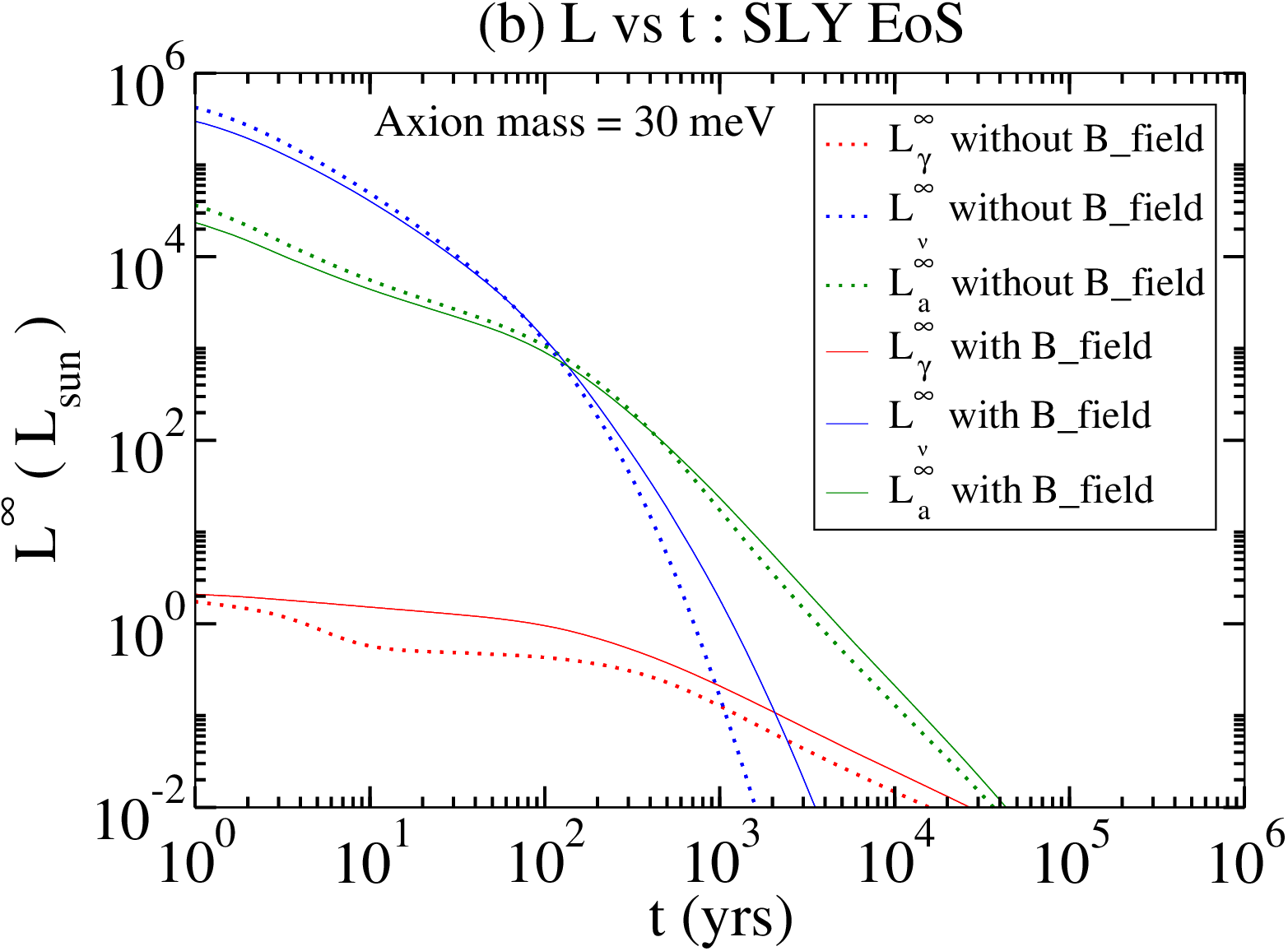}
\end{tabular}
\caption{ The variation of luminosity with the time of two different axion masses $1$ meV, \& $30$ meV, respectively, for SLY EoS in the presence and absence of magnetic field.}   
\label{fig:ltimevar7}
\end{figure*}

We have shown additional figures corresponding to axion mass 1 meV \& 30 meV.
In Figure (\ref{fig:ltimevar5}a) and Figure (\ref{fig:ltimevar5}b), we have presented the luminosity as a function of time at an axion masses $1$ meV \& $30$ meV for APR EoS respectively.
In Figure (\ref{fig:ltimevar6}a)\& Figure (\ref{fig:ltimevar6}b), we have presented the luminosity as a function of time at an axion masses $1$ meV \& $30$ meV for FPS EoS respectively.
In Figure (\ref{fig:ltimevar7}a)\& Figure (\ref{fig:ltimevar7}b), we have presented the luminosity as a function of time at an axion masses $1$ meV \& $30$ meV for SLY EoS respectively.

\end{document}